\documentclass[prb,aps, twocolumn, amsmath,amssymb,superscriptaddress]{revtex4}
\usepackage{float}
\usepackage{amsmath}
\usepackage{amssymb}
\usepackage{amsfonts}
\usepackage{euscript}
\usepackage{enumerate}
\usepackage{hhline}
\usepackage{pslatex}
\usepackage{tabularx}
\usepackage[usenames,dvipsnames]{xcolor}

\usepackage{graphicx}
\usepackage{dcolumn}
\usepackage{bm}
\usepackage[sort&compress]{natbib}

\makeatletter
\DeclareMathOperator*{\argmin}{arg\,min}

\newcommand{\erf}{\text{erf}}

\newcommand{\mum}{ \,\mu\text{m}}

\newcommand{\SiN}{\text{Si}_\text{3}\text{N}_\text{4}}

\renewcommand{\@biblabel}[1]{#1. }
\renewcommand{\@dotsep}{500}
\renewcommand{\@pnumwidth}{0em}
\renewcommand{\l@figure}[2]{
\@dottedtocline{1}{1.5em}{2em}{Figure #1}{}\vspace{15pt}}

\begin{document}

\title{A heterogeneous III-V/silicon integration platform for on-chip quantum photonic circuits with single quantum dot devices}

\author{Marcelo Davanco}\email{marcelo.davanco@nist.gov}
\affiliation{Center for Nanoscale Science and Technology, National
Institute of Standards and Technology, Gaithersburg, MD 20899,
USA}
\author{Jin Liu}\email{liujin23@mail.sysu.edu.cn}
\affiliation{Center for Nanoscale Science and Technology, National
Institute of Standards and Technology, Gaithersburg, MD 20899,
USA}
\affiliation{Maryland NanoCenter, University of Maryland, College Park, USA}
\affiliation{School of Physics, Sun-Yat Sen University, Guangzhou, 510275, China}

\author{Luca Sapienza}
\affiliation{Center for Nanoscale Science and Technology, National
Institute of Standards and Technology, Gaithersburg, MD 20899,
USA}
\affiliation{Department of Physics \& Astronomy, University of Southampton S017 1BJ, UK}

\author{Chen-Zhao Zhang}
\affiliation{South China Academy of Advanced Optoelectronics, Science Building No. 5, South China Normal University, Higher-Education Mega-Center, Guangzhou 510006, China}

\author{Jos\'e Vin\'icius De Miranda Cardoso}
\affiliation{Center for Nanoscale Science and Technology, National
Institute of Standards and Technology, Gaithersburg, MD 20899,
USA}
\affiliation{Federal University of Campina Grande, Brazil}

\author{Varun Verma}
\affiliation{National Institute of Standards and Technology, Boulder, CO 80305,
USA}

\author{Richard Mirin}
\affiliation{National Institute of Standards and Technology, Boulder, CO 80305,
USA}

\author{Sae Woo Nam}
\affiliation{National Institute of Standards and Technology, Boulder, CO 80305,
USA}

\author{Liu Liu}
\affiliation{South China Academy of Advanced Optoelectronics, Science Building No. 5, South China Normal University, Higher-Education Mega-Center, Guangzhou 510006, China}

\author{Kartik Srinivasan} \email{kartik.srinivasan@nist.gov}
\affiliation{Center for Nanoscale Science and Technology, National
Institute of Standards and Technology, Gaithersburg, MD 20899, USA}

\date{\today}

\begin{abstract}
\noindent \textbf{ Photonic integration is an enabling technology for photonic quantum science, offering greater scalability, stability, and functionality than traditional bulk optics. Here, we describe a scalable, heterogeneous III-V/silicon integration platform to produce $\SiN$ photonic circuits incorporating GaAs-based nanophotonic devices containing self-assembled InAs/GaAs quantum dots. We demonstrate pure single-photon emission from individual quantum dots in GaAs waveguides and cavities - where strong control of spontaneous emission rate is observed - directly launched into $\SiN$ waveguides with $>90~\%$ efficiency through evanescent coupling. To date, InAs/GaAs quantum dots constitute the most promising solid-state triggered single-photon sources, offering bright, pure and indistinguishable emission that can be electrically and optically controlled. $\SiN$ waveguides offer low-loss propagation, tailorable dispersion and high Kerr nonlinearities, desirable for linear and nonlinear optical signal processing down to the quantum level. We combine these two in an integration platform that will enable a new class of scalable, efficient and versatile integrated quantum photonic devices.}
\end{abstract}

 \maketitle

\maketitle

Although the increasing complexity of quantum photonic circuits has enabled small-scale demonstrations of quantum computation, simulations, and metrology~\cite{ref:Politi_OBrien,tanzilli_genesis_2012}, the development of highly-integrated systems that can solve more complex problems~\cite{ralph_quantum_2013} is severely limited by system inefficiencies. In circuits that are, by and large, composed of purely passive elements such as waveguide arrays, phase delays, and beamsplitters, a combination of small photon flux at the circuit input, passive losses in the circuit, and inefficient detection at the output leads to unrealistically long times for large-scale experiments~\cite{loredo_bosonsampling_2016}. While losses within the photonic circuit can be minimized through an appropriate choice of materials and waveguide architectures, and single-photon detection with almost 100~\% efficiency is now possible with superconducting single-photon detectors~\cite{marsili_detecting_2013}, the availability of a large on-chip single-photon flux is still a significant bottleneck. Overcoming such a limitation can enable not only further scaling of photonic quantum information experiments, but also quantum-level investigation of a variety of physical processes in nanophotonic and nanoplasmonic structures, such as Kerr nonlinearities~\cite{li_efficient_2016}, optomechanical interactions~\cite{riedinger_non-classical_2016} and single-photon nonlinearities~\cite{sun_quantum_2016,bennett_semiconductor_2016,maser_few-photon_2016,reinhard_strongly_2012,fushman_controlled_2008}.

\begin{center}
\begin{figure}
\begin{center}
\includegraphics[width=\linewidth]{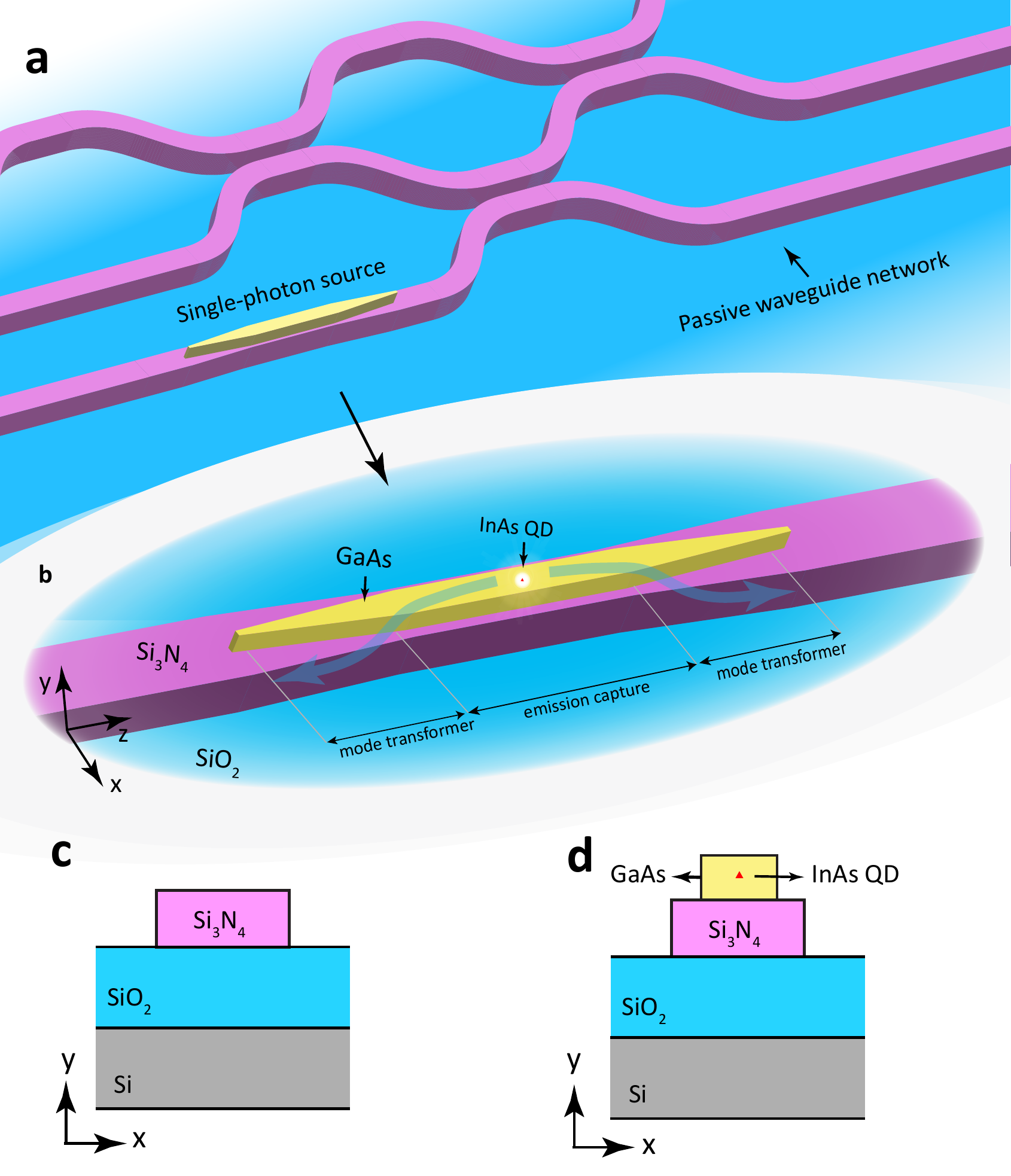}
\caption{\textbf{Principle of operation and device geometry}. \textbf{a}, Conceptual quantum photonic circuit composed of a passive waveguide network with a small, directly integrated nanophotonic single-photon source. \textbf{b}, Schematic showing geometry and operation principle of the integrated single-photon source shown in \textbf{a}. The source is composed of a single InAs QD embedded in a GaAs waveguide-based nanophotonic structure designed to efficiently capture QD emission and launch it directly into the $\SiN$ waveguide. \textbf{c} and \textbf{d}: Cross-sections of passive $\SiN$ and active GaAs waveguides that form the core elements of the integration platform.}
\label{fig:Fig1}
\end{center}
\end{figure}
\end{center}

The on-chip photon flux problem can be broken down into two components - low single-photon generation rates and low coupling efficiency from external sources into the chip. Single quantum dots (QDs) in microcavities have recently been shown to be capable of providing on-demand, pure, highly indistiguishable single-photons at high rates~\cite{somaschi_near-optimal_2016,ding_-demand_2016}, exceeding the performance of non-deterministic sources based on spontaneous nonlinear optical processes, and thus constitute a promising alternative to the source brightness issue. Nevertheless, coupling losses from free-space or optical fibers into the photonic circuit are still considerable. Although much effort is being devoted to improving the external source coupling efficiency, both in classical~\cite{cardenas_high_2014,notaros_ultra-efficient_2016} and quantum~\cite{ref:Davanco_WG,ref:Ates_Srinivasan_Sci_Rep} photonic experiments, it can be argued that such approaches, aimed at probing individual devices on a chip, do not allow for highly complex circuits. A scalable alternative is to instead create on-chip sources that are directly integrated into the passive waveguide network. This approach is in fact being pursued with spontaneous four-wave-mixing-type sources in silicon-based photonic circuits~\cite{silverstone_qubit_2015}, and in GaAs waveguide devices with QD on-chip sources~\cite{prtljaga_monolithic_2014,jons_monolithic_2015,reithmaier_-chip_2015,dietrich_gaas_2016}. In the former approach, which benefits from low-loss propagation in the silicon waveguides, the fundamental tradeoff between source brightness and purity limits the generated photon flux, and may require complex multiplexing schemes to produce quasi-deterministic sources~\cite{migdall_tailoring_2002,collins_integrated_2013}. In the latter case, which benefits from the availability of deterministic, high single-photon generation rates, propagation losses in the etched GaAs waveguides, considerably higher than in their silicon-based counterparts, limit on-chip photon routing, delay and interference; furthermore, the exclusive use of III-V materials in these architectures imposes challenging limits on device compactness, operation and performance (see SI for an extended discussion). Here, we present a photonic integration architecture that incorporates the benefits of these two approaches, and circumvents the aforementioned disadvantages. We have developed a scalable, integrated, heterogeneous III-V / silicon photonic platform to produce photonic circuits based on $\SiN$ waveguides that directly incorporate GaAs nanophotonic devices, such as waveguides, ring resonators, and photonic crystals, containing single self-assembled InAs/GaAs QDs. As illustrated in Figs.~\ref{fig:Fig1}a and~\ref{fig:Fig1}b, our integration platform allows the creation of passive, $\SiN$ waveguide-based circuits, which can be used for low-loss routing, distribution and interference of light across the chip. At select portions of such passive circuits, GaAs waveguide-based nanophotonic devices containing self-assembled InAs QDs are produced, on top of a $\SiN$ waveguide section. Such active GaAs devices can be designed to efficiently launch individual photons produced by the embedded QDs directly into the underlying $\SiN$ guide, acting as efficient on-chip triggered single-photon sources for the $\SiN$ waveguide circuit.

Self-assembled InAs/GaAs
QDs have been used to demonstrate close-to-optimal triggered single-photon emission~\cite{somaschi_near-optimal_2016,ding_-demand_2016}, spin-qubit operation~\cite{warburton_single_2013}, and a variety of strong-coupling cavity quantum electrodynamics (QED) systems~\cite{ref:Srinivasan16,sun_quantum_2016,fushman_controlled_2008}. The two-level system nature of these QDs, together with the ability to integrate them within nanophotonic devices, may ultimately form the basis of deterministic quantum gates. As a complementary technology, $\SiN$ waveguides offer low-loss propagation with tailorable dispersion and relatively high Kerr nonlinearities. These properties are currently being explored for linear ~\cite{xiong_compact_2015} and nonlinear~\cite{li_efficient_2016} optical signal processing, as well as cavity optomechanics-based measurements~\cite{purdy_observation_2016}, down to the quantum level.

Our work extends the application space of a mature, scalable, top-down heterogeneous photonic integrated circuit platform ~\cite{li_recent_2016} into the quantum realm. Other heterogeneous integration platforms for quantum photonics have been developed, in which silicon served only as a substrate and had no photonic function, through epitaxial growth of InAs/GaAs QDs on silicon~\cite{luxmoore_iiiv_2013}, or flip-chip bonding of QD-containing GaAs nanomembranes~\cite{chen_wavelength-tunable_2016}. Bottom-up or hybrid techniques~\cite{zadeh_deterministic_2016,mouradian_scalable_2015,murray_quantum_2015}, where nanostructures containing quantum emitters are produced separately and then transferred onto a photonic circuit have also been developed.  In contrast to all of these, our approach allows nearly independent, flexible, and high-resolution tailoring of both active (III-V) and passive (silicon) photonic waveguide elements with precise and repeatable, sub-100~nm alignment defined lithographically.  All of these characteristics meet the critical requirements for scalable integrated quantum photonic systems. Our platform is also amenable to electrical injection operation as shown in ref.~\onlinecite{khasminskaya_fully_2016}, along with all the aforementioned advantages offered by high performance single self-assembled QDs.

\noindent\textbf{Quantum dot interface design}
The schematic drawings in Fig.~\ref{fig:Fig1}c and Fig.~\ref{fig:Fig1}d respectively show cross-sections of passive and active waveguide sections that form the building blocks of our photonic integration platform. Passive sections consist of $\SiN$ ridges with SiO$_2$ and air for bottom and top claddings respectively. Active sections consist of the same $\SiN$ ridge, topped by a GaAs ridge containing a single InAs QD. Single-photon sources are created from active sections as indicated in Fig.~\ref{fig:Fig1}b.  The GaAs and $\SiN$ ridge widths are varied along $z$ over two separate sections, with complementary functions. The emission capture section collects light radiated by the QD into a guided wave confined to the GaAs ridge, and the mode transformer transfers light from the GaAs into the $\SiN$ ridge. We note that the source here is symmetric, so emission is in either $\pm z$ direction; unidirectional emission can be implemented with an an end mirror or through chiral coupling~\cite{coles_chirality_2016}. Design guidelines for the two sections are given in the following.

\begin{figure}
\includegraphics[width=\linewidth]{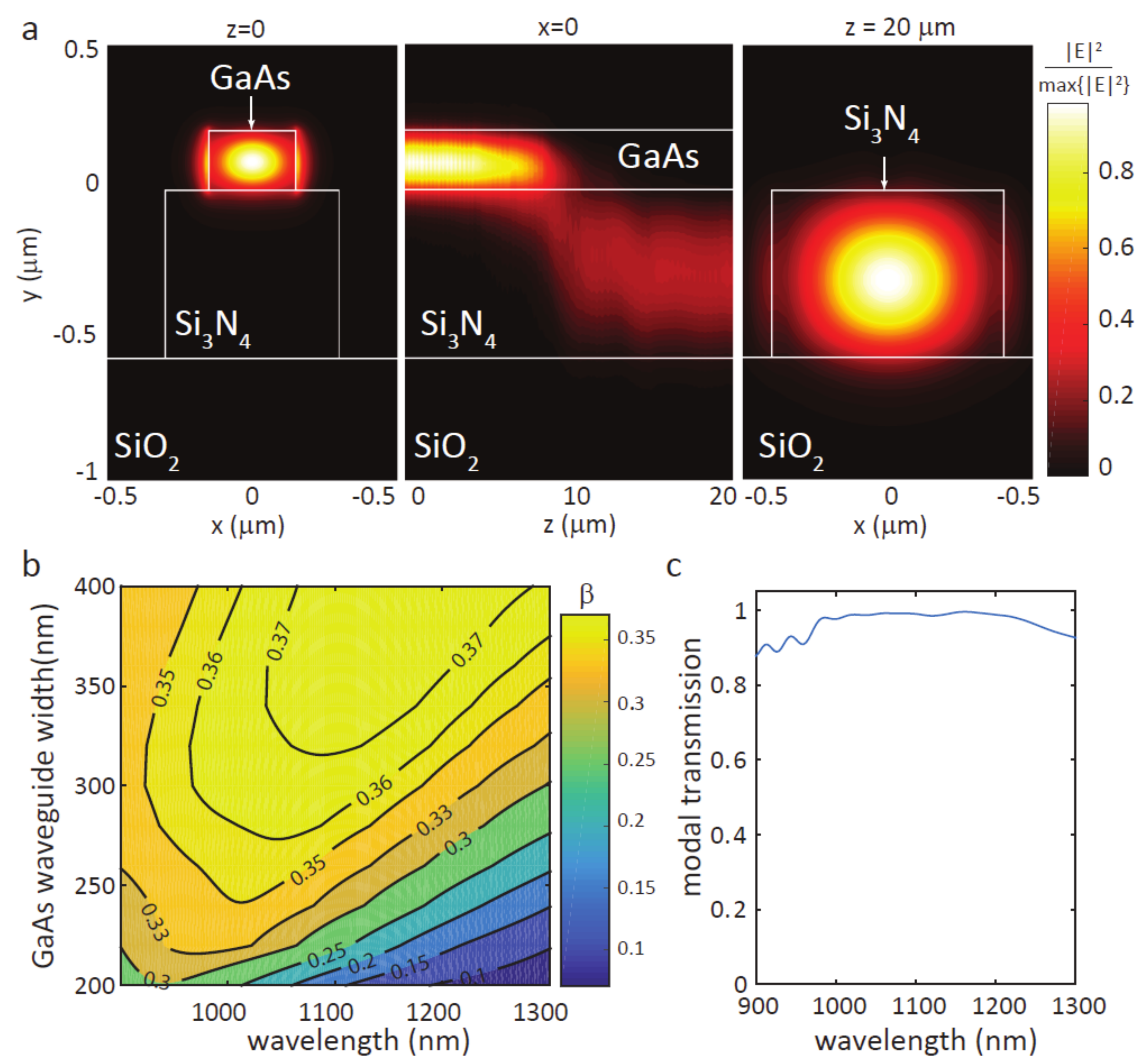}
\caption{\textbf{Nanophotonic design}. \textbf{a}, Left panel: Electric field distribution for the fundamental TE GaAs supermode of the waveguide stack in Fig.~\ref{fig:Fig1}d, with dimensions specified in the main text. Center panel: Electric field distribution across the mode-transformer cross-section, for a GaAs mode launched at $z=0$. At $z\approx10\mum$, the GaAs and $\SiN$ guides are phase-matched, and power is efficiently transferred from the top GaAs to the bottom $\SiN$ guide. Right panel: Fundamental TE mode of the $\SiN$ waveguide at the end of the mode transformer. \textbf{b}, Coupling efficiency ($\beta$), as a function of GaAs width and emission wavelength, of photons emitted by a dipole located at x=0 and 74~nm below the top surface, into the GaAs waveguide mode  traveling in either the +z or -z direction. \textbf{c}, Modal power conversion efficiency from the GaAs mode into the $\SiN$ mode in \textbf{a}, as a function of wavelength.}
\label{fig:Fig2}
\end{figure}

In the emission capture section, the widths of both waveguides are kept constant. The GaAs waveguide must support a single transverse-electric (TE) mode, and must be non-phase-matched to the $\SiN$ guide. This ensures that the fundamental TE supermode of the waveguide stack is strongly concentrated in the GaAs core, as shown in the left panel in Fig.~\ref{fig:Fig2}a. The InAs QD must then be made to radiate almost exclusively into the fundamental GaAs supermode, rather than into other guided or unbound modes of the stack. The ratio of the total dipole-emitted power that is coupled to the GaAs mode is the $\beta$-factor, $0\leq\beta\leq1$. $\beta\rightarrow 1$ can be achieved for guided modes in waveguides with high refractive index contrasts and sub-wavelength cross-sections, a result of strong field screening inside the guiding core, that takes place for radiative modes~\cite{ref:Bleuse2011}. This has been demonstrated in GaAs nanowires or nanowaveguides surrounded by air~\cite{ref:Claudon,ref:Davanco2,ref:Davanco_WG} or encapsulated in SiN~\cite{zadeh_deterministic_2016}. We predict similar performance for a GaAs nanowire on top of a $\SiN$ ridge. Assuming a horizontally ($x$) oriented QD electric dipole moment, we use finite difference time domain (FDTD) simulations to compute $\beta$ for the GaAs supermode of an active guide designed for emission wavelengths near 1100~nm. The thicknesses of the GaAs and $\SiN$ layers were taken from the wafer stack used for fabrication (see Methods and SI). Figure~\ref{fig:Fig2}b shows a contour map of $\beta$ as a function of wavelength and GaAs waveguide width, for a $\SiN$ waveguide thickness of 580~nm and width of 600~nm. For GaAs widths between 300~nm and 400~nm, $0.37>\beta>0.35$ for waves traveling in either the $+z$ or $-z$ direction ($0.74>2\cdot\beta>0.70$ total) is achievable over $\approx100$~nm around 1100~nm. Further simulations (not shown) indicate that $\beta$ is robust with respect to the $\SiN$ waveguide width, to within several tens of nm. Although $\beta$ is less than the maximum of 0.5 for symmetric emission, we note that both in simulations and in our devices the QD was located at a non-optimal vertical location inside the GaAs. In the SI, we provide similar simulations for an optimized geometry with $\beta > 0.45$ ($2\beta > 0.9$), comparable to those predicted in GaAs nanowires and nanowaveguides~\cite{ref:Claudon,ref:Davanco2,ref:Davanco_WG}, and in photonic crystal (PhC) slow-light waveguides~\cite{manga_rao_single_2007,lund-hansen_experimental_2008}. We note nevertheless that the capture section can be replaced by any type of waveguide-based device, such as PhC or microring resonators (demonstrated below), which may provide high $\beta$ through Purcell enhancement.

The mode transformer section consists of an adiabatic structure in which the widths of the GaAs and $\SiN$ waveguides are, respectively, reduced and increased along the $z$-direction. The width tapers are designed such that the two waveguides become phase-matched over some finite length along the mode converter, where power is efficiently transferred from the GaAs to the $\SiN$ guide; past the phase-matching length, the taper brings the two guides again away from the phase-matching condition, preventing the power from returning to the top guide. This is illustrated in the middle panel of Fig.~\ref{fig:Fig2}a, which shows the FDTD simulated electric field distribution for a transformer in which the GaAs and $\SiN$ widths vary linearly from 300~nm to 100~nm and from 800~nm to 600~nm respectively, over a length of 20~$\mu$m. Significantly shorter lengths can potentially be achieved with more sophisticated profiles~\cite{xia_photonic_2005}. Figure~\ref{fig:Fig2}c shows modal power conversion efficiency from the GaAs mode to the $\SiN$ mode (right panel of Fig.~\ref{fig:Fig2}a) as function of wavelength (see Methods for simulation details). Maximum efficiency in excess of 98~$\%$ is achieved over a $>200$~nm wavelength range. The geometry is robust to variations of tens of nm in the initial and final widths, well within electron-beam lithography tolerances.

With these two elements, the maximum efficiency of our ideal single-photon source is $\beta\cdot\eta\approx0.72$ into both directions of the $\SiN$ waveguide, or $36~\%$ in either the $+z$ or $-z$ direction. For the optimized design in the SI, efficiency $>90$~\% could potentially be achieved. We now describe the fabrication of the devices just discussed.

\begin{figure*}
\begin{center}
\includegraphics[width=\linewidth]{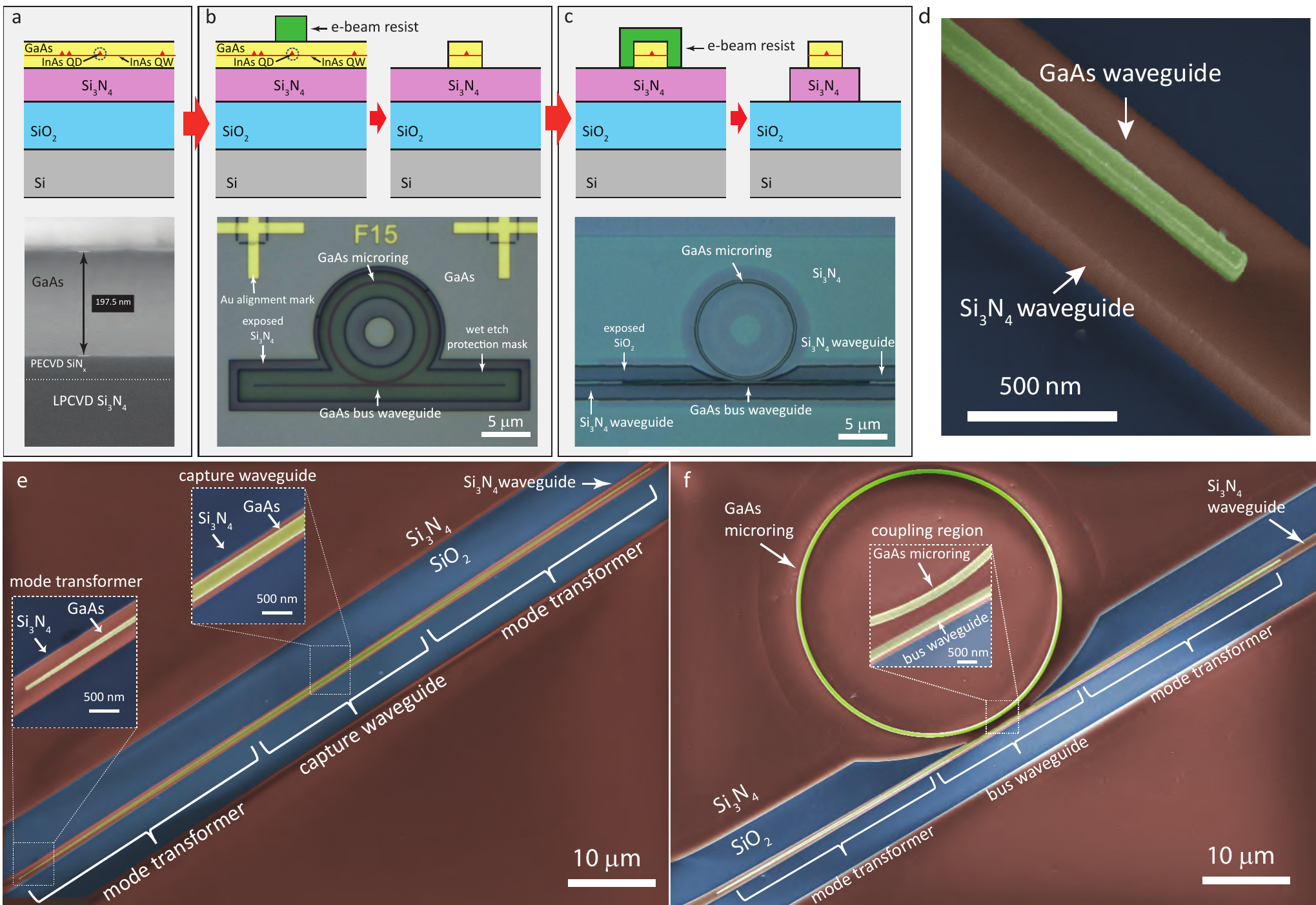}
\caption{\textbf{Device fabrication}. \textbf{a}, Top: schematic of bonded wafer stack used in fabrication, consisting of a top III-V layer, containing InAs QDs, that is directly bonded on top of a Si / 3$~\mum$ SiO$_2$/ 550~nm $\SiN$ stack. Bottom: cross-sectional scanning electron microgaph (SEM) of bonded wafer stack. The $\approx30$~nm SiN$_x$ layer was grown on the GaAs wafer surface prior to bonding (see SI details). \textbf{b}, Top: GaAs device lithography and etching steps. Bottom: optical micrograph of etched GaAs microring resonator and bus waveguide. Au aligment marks used for registered electron-beam lithography are visible. A wet etch protection resist mask (not depicted in the schematic - see SI) is also visible. \textbf{c}, Top: $\SiN$ waveguide lithography (aligned to the previously etched GaAs device) and etching steps. Bottom: optical micrograph of GaAs microring resonator and bus waveguide, and underlying $\SiN$ waveguide. \textbf{d}, False-color SEM of tip of mode-transformer geometry, common to both devices in \textbf{e} and \textbf{f}. \textbf{e}, False-color SEM of fabricated GaAs waveguide (yellow) on top of $\SiN$ (red) waveguide. Blue regions are exposed SiO$_2$. Insets show details of the mode transformer end tip and the QD photon capture waveguide. \textbf{f}, False-color SEM of GaAs microring and bus waveguide, and underlying $\SiN$ waveguide. Inset shows details of the microring-bus waveguide evanescent coupling region.}
\label{fig:Fig3}
\end{center}
\end{figure*}

\noindent\textbf{Heterogenous device integration}
We start with the wafer stack shown in Fig.~\ref{fig:Fig3}a. It consists of a silicon substrate topped by a 3~$\mum$ thick thermal oxide layer, a 550~nm layer of stoichiometric $\SiN$, and an epitaxially grown 200~nm GaAs/AlGaAs stack containing a single layer of InAs quantum dots-in-a-well (DWELL)~\cite{stintz_low-threshold_2000} located 74~nm below the top GaAs surface (details in the SI). As a result of the self-assembled growth, quantum dots were randomly distributed within this layer, with a density $>100/$~$\mum^2$. The hybrid III-V semiconductor / $\SiN$ stack is produced with a low-temperature, oxygen plasma-activated wafer bonding procedure~\cite{li_recent_2016} detailed in the SI. Following the wafer bonding step, fabrication proceeds as in Figs.~\ref{fig:Fig3}b and ~\ref{fig:Fig3}c (optical micrographs of the devices after completion of each step are also shown). An array of Au alignment marks is first produced on top of the GaAs layer via electron-beam lithography followed by metal lift-off. Electron-beam lithography and inductively-coupled plasma etching are next used to define GaAs devices aligned to the Au mark array. After cleanup of the etched sample surface, electron-beam lithography referenced to the same Au mark array is performed to define $\SiN$ waveguide patterns aligned to the previously etched GaAs devices. Reactive ion etching is then used to produce the $\SiN$ waveguides. As a final step, the chip is cleaved perpendicularly to the $\SiN$ waveguides  $>1$~mm away from the GaAs devices, to allow access with optical fibers in the endfire configuration. Before cleaving, 168 devices were produced, with a $>80~\%$ overall yield considering just device geometry. Features as small as 50~nm were achieved in the GaAs layer, and alignment accuracy on the order of a few tens of nm between the top and bottom waveguides was typically observed. We point out that, although here we had no control over QD location within the fabricated GaAs devices, we have specifically tailored our fabrication sequence to allow seamless incorporation of positioning techniques capable of spatially mapping QDs with respect to the Au marks~\cite{sapienza_nanoscale_2015,dousse_controlled_2008}.

Figure~\ref{fig:Fig3}d is a false-color scanning electron micrograph (SEM) of a fabricated stacked-waveguide structure, corresponding to the tip of a mode transformer section. GaAs, $\SiN$ and SiO$_2$ are colored in yellow, pink and blue respectively. Figures~\ref{fig:Fig3}e and ~\ref{fig:Fig3}f show SEMs of two types of fabricated devices, with different emission capture geometries. In Fig.~\ref{fig:Fig3}e, the capture structure is a straight waveguide as discussed above. The insets show details of the capture and mode transformer sections. In Fig.~\ref{fig:Fig3}f, the capture structure is a GaAs microring resonator that is evanescently coupled to a bus waveguide with mode transformers, with the same geometry as in Fig.~\ref{fig:Fig3}e. Here, QD emission coupled to whispering-gallery modes of the GaAs microring are outcoupled through the bus waveguide (coupling region shown in the inset), and then transferred to the $\SiN$ guide via the mode transformers. We next describe optical measurements done to characterize the photonic performance of the fabricated devices.

\noindent\textbf{Mode transformer characterization}
Two important parameters common to all types of devices are the mode transformer efficiency $\eta$ and the external coupling efficiency $\eta_\text{ ext}$. The first determines, together with the $\beta$-factor, the efficiency of the interface between the QD-containing GaAs layer and the passive waveguide circuit. The latter is the efficiency with which the device can be accessed from off-chip, ultimately determining the absolute power available for detection.

\begin{center}
\begin{figure*}
\begin{center}
\includegraphics[width=\linewidth]{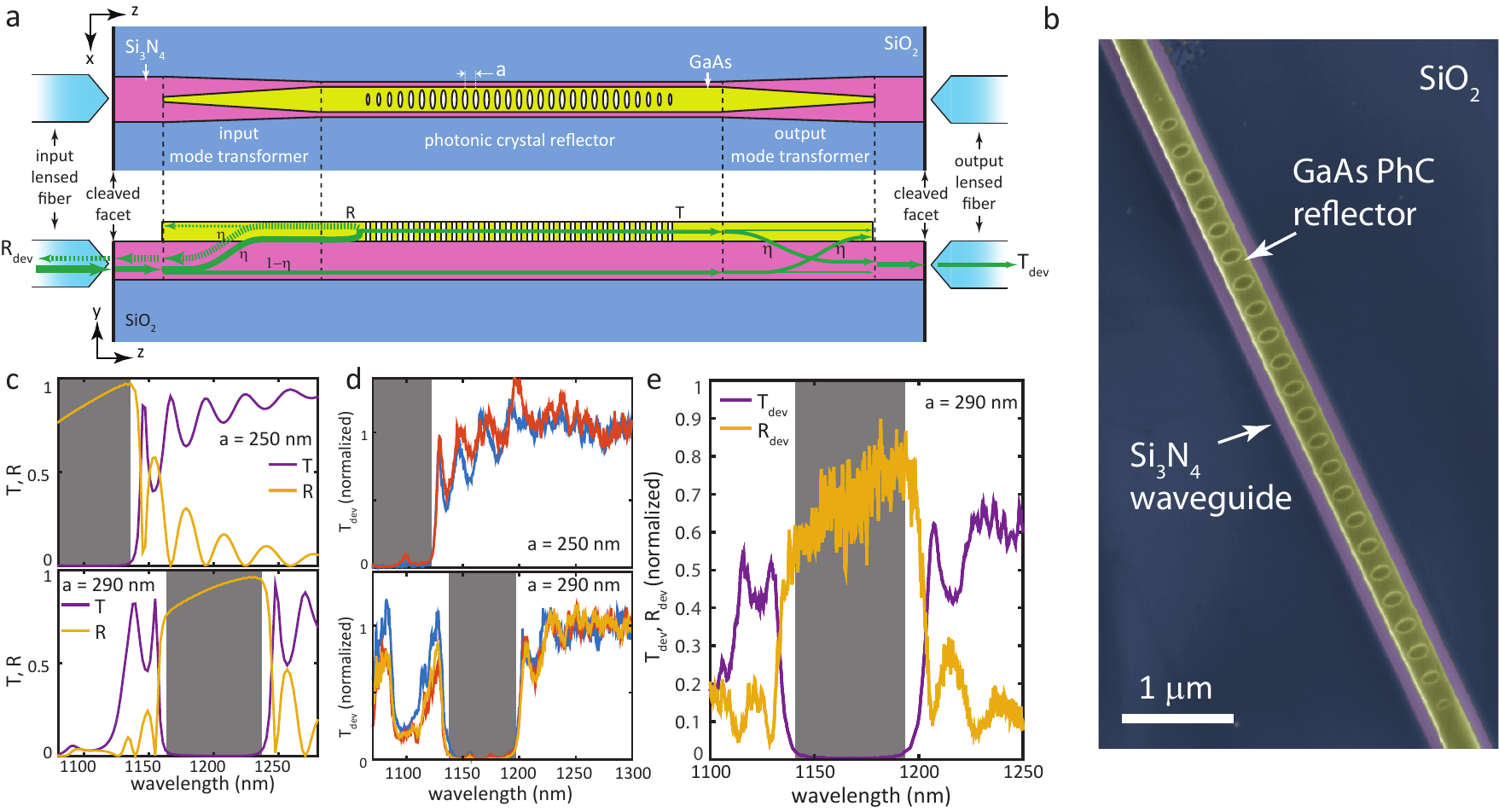}
\caption{\textbf{Characterizing mode transformer efficiency with a photonic crystal reflector.} \textbf{a}, Schematic of a PhC reflector device. Top: top-view. Bottom: cross-section. Green arrows indicate pathways taken by the optical signal injected at the input port. R and T stand for PhC modal power transmission and reflection spectra, and T$_\text{dev}$ R$_\text{dev}$ transmission and reflection spectra through the entire device, including lensed fibers. \textbf{b}, False-color SEM of fabricated GaAs PhC reflector (yellow) on top of a $\SiN$ (pink) waveguide, on top of exposed SiO$_2$ (blue). \textbf{c}, FDTD-simulated TE modal transmission (T, purple) and reflection (R,yellow) spectra as a function of wavelength for the PhC (without mode transformers), for two different lattice constants $a$. \textbf{d}, Experimental transmission spectra for various PhC reflectors with $a=250$~nm (top) and $a=290$~nm (bottom), normalized first by the transmission spectrum of a baseline $\SiN$ waveguide (without GaAs sections), then to the mean transmission at wavelengths between 1250~nm and 1300 nm. Different colors indicate different devices. \textbf{e}, Experimental transmission and reflection spectra for a PhC reflector with $a=290$~nm, normalized to the transmission spectrum of a baseline $\SiN$ waveguide (SI for details). Grey areas have transmission $<-15$~dB in \textbf{c}-\textbf{d}, $<-20$~dB in \textbf{e}.}
\label{fig:Fig4}
\end{center}
\end{figure*}
\end{center}

We estimate the mode transformer $\eta$ via transmission spectroscopy of a third type of device we fabricate within our platform, a waveguide-coupled photonic crystal (PhC) reflector. A schematic of the device in shown in Fig.~\ref{fig:Fig4}a. The PhC is a $\approx300$~nm wide GaAs waveguide into which a periodic 1D array of elliptical holes is etched, with lattice constant $a$. Major and minor hole radii are kept constant over 19 lattice constants at the center, then reduced linearly over 5 constants at the two ends of the array (to minimize radiation losses). The false-color SEM in Fig.~\ref{fig:Fig4}b illustrates the type of high resolution GaAs devices achievable within our platform. The periodic hole array defines a photonic bandgap for the TE-polarized GaAs mode on the left panel of Fig.~\ref{fig:Fig2}a. The latter is strongly reflected by the PhC at bandgap wavelengths. Figure~\ref{fig:Fig4}a describes the PhC reflector operation. Light is launched into the $\SiN$ waveguide using a lensed optical fiber aligned to its cleaved facet, then transferred with efficiency $\eta$ to the GaAs waveguide via the input mode transformer. At bandgap wavelengths, the GaAs-guided light is reflected with reflectivity $R$ by the PhC, then transferred back into the $\SiN$ waveguide via the input transformer, with efficiency $\eta$.

Simulated TE GaAs mode power transmission (T) and reflection (R) spectra are shown in Fig.~\ref{fig:Fig4}c, for PhCs with $a=250$~nm and $a=290$~nm and dimensions estimated by SEM from fabricated devices. Photonic bandgaps are evidenced by high reflectivity, high transmission extinction spectral regions marked in grey. We emphasize that R and T are spectra for the GaAs-confined modes, i.e., they do not include effects due to the mode transformers. We nevertheless observe, in Fig.~\ref{fig:Fig4}d, similar features experimentally, which suggests spectrally broad mode transformer operation consistent with Fig.~\ref{fig:Fig2}c. The experimental setup used is described in the Methods and SI. Room-temperature characterization is adequate to assess the low-temperature performance given the spectrally broadband nature of the elements involved and the expected thermo-optic shift of GaAs. Figure~\ref{fig:Fig4}d shows normalized experimental TE-polarized transmission spectra for various fabricated devices with either $a=250$~nm or $a=290$~nm. Consistent spectral features achieved across many devices indicate that our photonic integration platform is scalable. Figure~\ref{fig:Fig4}e shows a typical PhC reflectivity ($R_\text{ dev}$) peak, obtained for one of the $a=290$~nm devices, spectrally aligned with the transmission extinction region. The $>20$~dB ($\approx25$~dB at bandgap center) extinction highlighted in grey indicates highly efficient coupling from the $\SiN$ access waveguide into the GaAs layer, since light not transferred to the GaAs is not reflected by the PhC. As described in the Methods, the photonic bandgap extinction can be used to obtain a lower bound for the mode transformer efficiency $\eta$. For a typically observed 20~dB extinction, $\eta>90$~\%, conservatively. For the peak extinction of $\approx25$~dB, $\eta>94$~\%.

To determine the external coupling efficiency $\eta_\text{ ext}$, we took the transmitted power spectrum of a blank $\SiN$ waveguide (i.e., with no GaAs devices) and normalized it by the supercontinuum source power spectrum. Assuming identical waveguide facets on both chip edges, $\eta_\text{ ext}=0.23~\pm~0.03$ over the 1100~nm to 1300~nm wavelength range, across three different devices (uncertainties are propagated single standard deviations. See SI for transmission spectra). To verify this, we estimated a mode-mismatch coupling efficiency $\eta_\text{facet}\approx26$~\% between the $\SiN$ waveguide mode and a Gaussian beam with $2.5\mum$ diameter, consistent with the nominal lensed fiber spot-size diameter. The small difference between the experimental coupling efficiency and the calculated value suggests that propagation losses in the waveguide are relatively small.
\noindent\textbf{Quantum dot coupling to waveguides in the heterogeneous platform}
We next investigated QD emission coupling in our devices via photoluminescence (PL) measurements at cryogenic temperatures. In our setup, shown in Fig.S2 in the SI, devices were placed inside a liquid Helium flow cryostat, kept fixed on a copper mount connected to the cold finger. Testing temperatures ranged between 7~K and 30~K. A microscope system allowed individual devices to be visually located and optically pumped with laser light focused through a microscope objective. PL was collected by aligning a lensed fiber (mounted on a xyz nanopositioning stage inside the cryostat) to the corresponding $\SiN$ waveguide facet. The collected PL was either sent to a grating spectrometer equipped with a liquid nitrogen cooled InGaAs detector array for spectrum measurements, or towards a pair of amorphous WSi superconducting nanowire single-photon detectors (SNSPDs)~\cite{marsili_detecting_2013} for time-correlated single photon counting (TCSPC) measurements. We note that the high density QD population in our sample displayed a wide inhomogeneously broadened spectrum, with ensemble s-shell and p-shell peaks located approximately at 1100~nm and 1060~nm respectively.

\begin{center}
\begin{figure}
\begin{center}
\includegraphics[width=\linewidth]{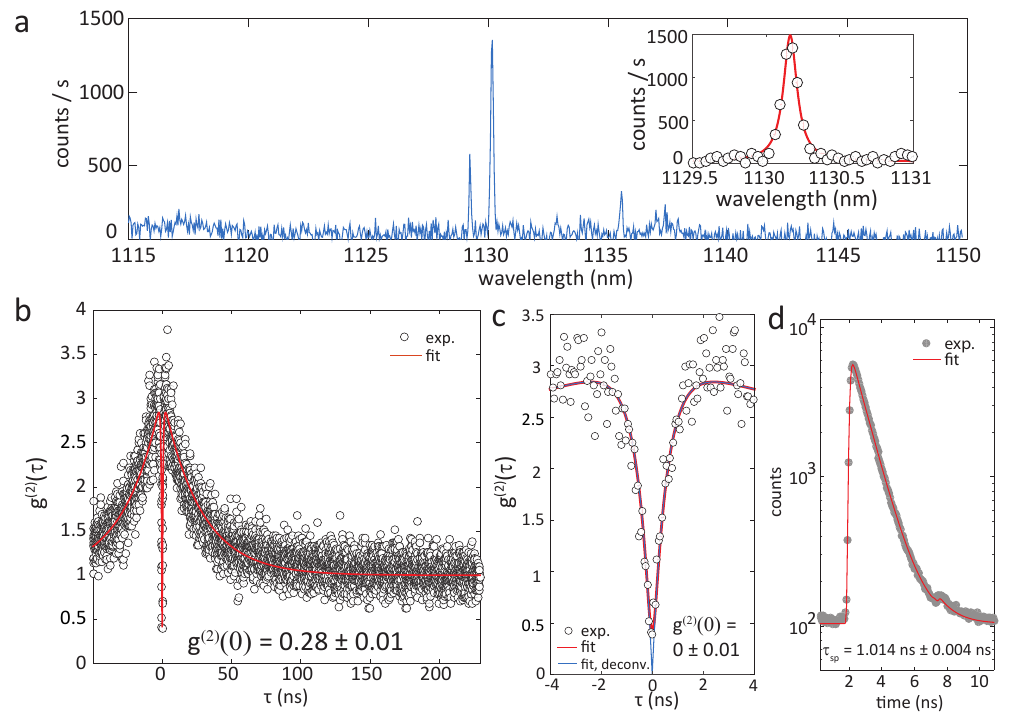}
\caption{\textbf{Quantum dot-waveguide coupling}. \textbf{a}, Photoluminescence spectrum for a single QD inside a GaAs waveguide as in Fig.~\ref{fig:Fig3}\textbf{e}, pumped with 1061~nm wavelength laser light. The PL is transferred to the bottom $\SiN$ waveguide, and collected with a lensed optical fiber inside of a Liquid Helium flow cryostat (see SI for experimental details). Sharp lines are exciton transitions from a single QD. Inset: Fit of PL peak at 1130.18~nm. \textbf{b}, Second-order correlation as a function of time delay $\tau$ for the 1130.18~nm line. Circles mark experimental data, red line is a fit (see Methods and SI). \textbf{c}, Zoom-in of \textbf{b} near $\tau=0$. The blue curve and quoted $g^{(2)}(0)$ are obtained from the red fit by deconvolving the detection time-response. Uncertainties for $g^{(2)}(0)$ are 95~\% fit confidence intervals (two standard deviations). \textbf{d}, Photoluminescence decay trace for the 1130.18~nm line. Gray dots are experimental data, the red line is a fit with a monoexponential function with lifetime $\tau_\text{sp}$. The uncertainty is obtained from the fit and corresponds to a single standard deviation.}
\label{fig:Fig5}
\end{center}
\end{figure}
\end{center}

We first investigated QD emission inside the basic hybrid device, a $\approx300$~nm wide, 10~$\mum$ long GaAs waveguide with $20~\mum$ long mode transformers, coupled to a 800 nm wide $\SiN$ waveguide. Figure~\ref{fig:Fig5}a shows the PL spectrum collected at a temperature of $\approx7$~K for a device pumped at $\lambda=1061$~nm (p-shell) with an tunable external-cavity diode laser (ECDL). Sharp spectral lines are excitonic complexes of individual QDs. A $\approx700$~pm full-width at half-maximum (FWHM) bandpass grating filter was used to spectrally isolate the line at 1130.18~nm in Fig.~\ref{fig:Fig5}a, and a Hanbury-Brown and Twiss (HBT) setup was used to measure the autocorrelation $g^{(2)}(\tau)$, in Figs.~\ref{fig:Fig5}b and ~\ref{fig:Fig5}c. The values $g^{(2)}(0)= 0.41 \pm 0.13$ obtained for the raw data, and $g^{(2)}(0) = 0\pm 0.13$ obtained by taking into account the $\approx129$~ps time resolution of our TCSPC system (see Methods), indicate that the QD in the GaAs device acts as source of single-photons that are directly launched into a $\SiN$ waveguide. $g^{(2)}(0)$ uncertainties quoted here and below are $95~\%$ fit confidence intervals (two standard deviations). Bunching at $\tau\approx\pm2$~ns suggests QD blinking as observed with quasi-resonat (p-shell) excitation in ref.~\onlinecite{santori_submicrosecond_2004}, and could be related to coupling of the radiative excited state to dark states. Our fits were done with a function that models coupling of a two-level system to a single dark state~\cite{davanco_multiple_2014}.

Lifetime measurements for the same QD line were next performed by modulating the ECDL pump light with an electro-optic modulator (see Methods and SI). The decay curves show in Fig.~\ref{fig:Fig5}d were fitted with a single exponential function, revealing a lifetime $\tau_\text{sp} = 1.014$~ns$~\pm~0.004$~ns (lifetime uncertainties here and below are from the fit and correspond to one standard deviation).  Assuming a fiber-to-chip coupling efficiency of $22~\%$, and a coupler efficiency $\eta=98$~\%, we estimate a QD-waveguide coupling parameter $\beta=0.20 \pm 0.07$ (uncertainty from propagated errors in the optical characterization of the measurement system, corresponding to one standard deviation. See Methods for details). This value, though appreciable, is less than the theoretical maximum of 0.37. This discrepancy could be attributed to non-optimal QD position and electric dipole moment orientation.

\begin{center}
\begin{figure*}
\begin{center}
\includegraphics[width=\linewidth]{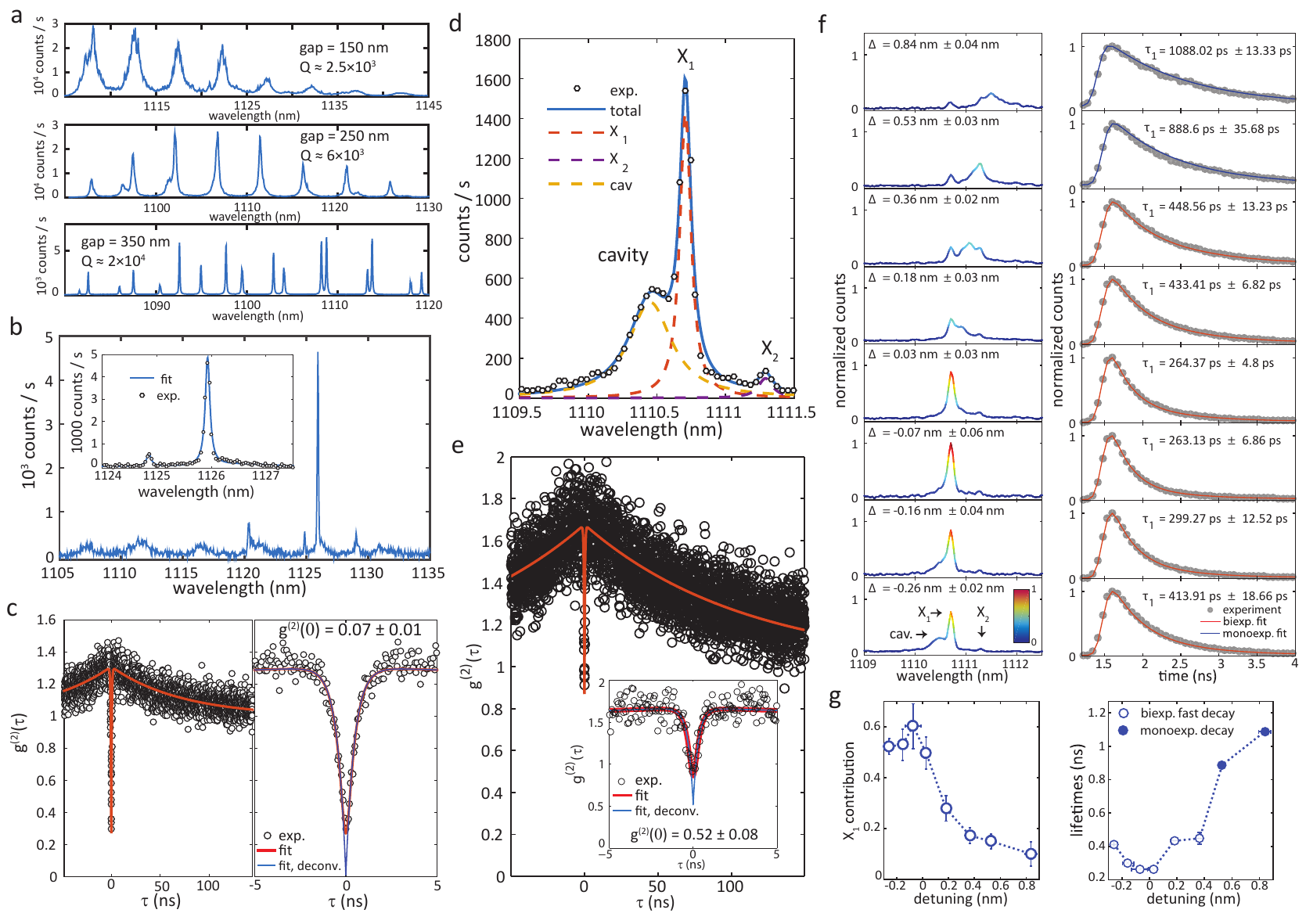}
\caption{\textbf{Quantum dot-cavity coupling}.\textbf{a}, Photoluminescence (PL) spectra as a function of wavelength from a QD ensemble pumped with laser light at 975~nm, emitting inside three different GaAs microring resonators. Peaks are whispering-gallery modes (WGMs) with quality factor Q, which increases with the microring-bus waveguide gap width. \textbf{b}, PL spectrum for a single QD coupled to a $Q\approx1.1\times10^4$ WGM. Inset: fit of cavity-coupled QD emission near 1126~nm. \textbf{c}, Left: second-order correlation $g^{2}(\tau)$ for the 1126~nm exciton line in \textbf{b}. Right: close-up near $\tau=0$. Circles are experimental data, red lines are a fit. The blue curve and quoted $g^{(2)}(0)$ are obtained from the red fit by deconvolving the detection time-response. \textbf{d}, PL spectrum for a single QD in a microring, coupled to a $Q\approx2.3\times10^4$ WGM. Circles: experimental data. Blue continuous line: fit. Dashed lines: fitting Lorentzians for the cavity and two excitons, X$_1$ and X$_2$. \textbf{e}, $g^{2}(\tau)$ for X$_1$ in \textbf{d}. Inset: close-up near $\tau=0$. \textbf{f}, Left panel: PL spectra for varying spectral detuning $\Delta$ between X$_1$ and the cavity. $\Delta$ is obtained from fits as in~\textbf{d}. All spectra are normalized to the intensity maximum at $\Delta\approx-0.07$~nm. The color scale indicates normalized intensity. Right panel: corresponding $X_1$ photoluminescence decay curves. Grey dots are experimental data, red (blue) lines are biexponential (monoexponential) decay fits. For biexponential fits, $\tau_1$ is the fast lifetime. \textbf{g}, Left panel: integrated intensity as a function of $\Delta$ for the filtered X$_1$ exciton contribution to the the PL spectra in \textbf{f}, obtained from Lorentzian fits as in \textbf{d}, normalized by the integrated intensity of the full fitted spectrum. Right panel: decay lifetimes for the fits in \textbf{f}, as a function of $\Delta$. Open blue circles are the fast biexponential decay lifetimes, closed blue circles are the monoexponential decay lifetimes. Uncertainties for $g^{(2)}(0)$, $\Delta$ and the X$_1$ magnitude are 95~\% fit confidence intervals (two standard deviations). Lifetime uncertainties are single standard deviations from the exponential decay fits.}
\label{fig:Fig6}
\end{center}
\end{figure*}
\end{center}

\noindent\textbf{Weak-coupling cavity QED in the heterogeneous platform}
We next investigated cavity effects on the radiative rate of single QDs coupled to whispering gallery modes (WGMs) of GaAs microring resonators (Fig.~\ref{fig:Fig3}f). The devices consisted of 20-$\mum$ diameter microrings formed by $\approx300$~nm wide waveguides, evanescently coupled to $\approx300$~nm wide GaAs bus waveguides spaced by gaps of varying dimensions. In this scheme, light from QDs inside the ring is outcoupled through the bus waveguide and then transferred to the $\SiN$ waveguide via the mode transformers. Figure~\ref{fig:Fig6}a shows PL spectra for three different resonators, with coupling gaps of 150~nm, 250~nm and 350~nm, pumped at high intensities with 975~nm laser light (resonant with the quantum well transitions). Peaks are PL from the QD ensemble coupled to WGMs. Quality factors for devices with the gap spacings of 150~nm, 250~nm and 350~nm are $2.5\times10^3$, $6\times10^3$ and $2\times10^4$. The increased Q for larger gaps is due to a decreased cavity-bus waveguide coupling, indicating that the geometrical control afforded by our fabrication platform enables fine control of cavity outcoupling rates. Pumping one of the $Q\approx1.1\times10^4$ microresonators at 1058~nm (p-shell) allowed observation of the single QD excitonic line at 1125.92~nm in Fig.~\ref{fig:Fig6}b, which was coupled to one of the cavity's WGMs. Background emission, likely from other QDs and (multi)excitonic complexes in the active material, is also observed in the different WGMs. Figure~\ref{fig:Fig6}c indicates the cavity-coupled QD acts as a single-photon source with $g^{(2)}(0)=0.28~\pm~0.01$ ($g^{(2)}(0)=0.07~\pm~0.01$ adjusted for detection time resolution).

We next demonstrated tunable control of Purcell radiative rate enhancement in a device with $Q\approx2.3\times10^4$, at a fixed temperature of $\approx$7~K. Pumping at $\lambda=1065$~nm (p-shell) allowed us to observe the cavity-mode-coupled single QD exciton line $X_1$ in Fig.~\ref{fig:Fig6}d, as well as a cavity-detuned exciton $X_2$. For the $X_1$ line, as seen in Fig.~\ref{fig:Fig6}e, $g^{(2)}(0)=0.72\pm0.08>0.5$ ($g^{(2)}(0)=0.52\pm0.08$ adjusted for detection time resolution), due to background emission from the cavity mode, which was transmitted by the band-pass filter introduced before detection. Indeed, based on the fit shown in  Fig.~\ref{fig:Fig6}d, cavity emission corresponds to $\approx45~\%$ of the filtered light intensity. To tune the cavity with respect to the QD exciton, we used the nitrogen gas-tuning mechanism of ref.~\onlinecite{srinivasan_optical_2007}. A small amount of gaseous $N_2$ is introduced in steps into the cryostat, and gettering at the GaAs surfaces red-shifts the cavity resonance by a small amount at each step. This is observed in the left panel in Fig.~\ref{fig:Fig6}f, where the PL spectrum of the cavity-coupled QD exciton ($X_1$) is seen to grow in intensity as its spectral (wavelength) detuning $\Delta$ from the cavity center tends to zero. The variation in intensity comes together with a variation in the exciton lifetime, evident in the corresponding decay curves on the right panel of Fig.~\ref{fig:Fig6}f. Biexponential fits to the decay data (monoexponential for $\Delta\approx0.53$~nm and $\Delta\approx0.84$~nm ) are also shown. The detuning-dependent variations in $X_1$ intensity and decay lifetime are summarized respectively in the left and right panels in Fig.~\ref{fig:Fig6}g, evidencing high-resolution, strong control of the exciton radiative rate via cavity coupling achieved in our platform. Further details on PL spectrum and decay fitting and assignment of lifetimes are given in the SI.
Comparing with the $\approx$1~ns lifetime in the waveguide, we can extract a maximum radiative rate enhancement factor of $\approx4$ for the QD. From the calculated WGM mode volume $V_{eff}=75.5(\lambda/n_\text{GaAs})^3$ ($n_\text{GaAs}$ is the GaAs refractive index) and the experimental $Q=2.3\times10^4$, we expect a maximum Purcell Factor $F_p\approx23$ (see Methods). The lower Purcell factor observed in experiment could be due to non-optimal spatial location and polarization alignment of the QD with respect to the microring mode.

\noindent\textbf{Discussion}
The results presented demonstrate that our platform enables the creation of integrated photonic circuits that incorporate quantum-dot based devices with complex geometries. As discussed above and in the SI, further improvements to the single-photon capture efficiency (quantified by the $\beta$-factor) can be achieved through optimized wafer stacks (both $\SiN$ and the GaAs epi-stack) and device geometries. In particular, our platform allows the creation of geometries providing high Purcell radiative rate enhancement where high $\beta$ may be achieved, such as microdisk, microcring or photonic crystal-based cavities and slow-light waveguides. The high reflectivity achieved with our PhC reflectors furthermore suggests a path forward towards unidirectional QD emission in a waveguide. Alternatively, chiral coupling to waveguide modes~\cite{coles_chirality_2016} could also be explored. Strongly-coupled QD-cavity systems~\cite{ref:Srinivasan16,sun_quantum_2016,fushman_controlled_2008} evanescently coupled to a bus waveguide could also be envisioned in our platform.

As mentioned above, our III-V wafers contained a high density of QDs ($>100/\mum^2$), randomly distributed across the wafer surface, which led to the deterioration of the purity of our on-chip single-photon sources. It is also possible that the pronounced blinking observed in the autocorrelation traces might stem from interactions between many neighboring QDs. Low-density QD growth constitutes a clear way forward here. In this case, QD positioning techniques such as the one developed in ref.~\onlinecite{sapienza_nanoscale_2015} -a technique fully compatible with our fabrication process- become essential. Precise quantum dot location within a nanophotonic structure would also allow $\beta$ and Purcell factor optimization.

The underlying $\SiN$ waveguides demonstrated here provide not only a way to route single-photons with low loss across the chip, but also a means to explore nonlinear optical processes with single photons. For instance, four-wave-mixing-based wavelength conversion of single-photon-level laser light was recently demonstrated in a $\SiN$ microring resonator with cross-sectional dimensions similar to those of our waveguides, and fabricated with the same etch process~\cite{li_efficient_2016}. This means that the required dispersion profiles and nonlinear coefficients are attainable within our platform. At the same time, passive structures with cross-sections optimized for low propagation losses may also be implemented, for instance with thinner $\SiN$ (see SI) and potentially even with a top oxide cladding. The introduction of elements such as on-chip delay lines, high quality $\SiN$-based filters, and microring add-drops, can also be envisioned.

Our platform is also amenable to further integration with waveguide-based superconducting nanowire single-photon detectors~\cite{pernice_high-speed_2012}. Finally, the fabrication process can be adapted for materials such as AlN and LiNbO$_3$, which may enable active electro-optic phase control. We anticipate all of these features will enable a new class of monolithic on-chip devices comprising emission, routing, modulation and detection of quantum light.

\noindent \large{\textbf{Methods}}

\noindent \small{\textbf{Numerical simulation}}
Calculations of waveguide $\beta$-factors is done with finite-difference time-domain simulations. We simulate a $x$-oriented electric dipole source radiating inside the GaAs ridge of the stacked GaAs/$\SiN$ waveguide structure shown in Fig.~\ref{fig:Fig1}d. The simulation is 3D, and the coupled waveguide structure length is $1~\mum$. Perfectly-matched layers are used to emulate either open regions (air and SiO$_2$ semi-infinite spaces above and below the geometry), or infinite waveguides (in the planes perpendicular to $x$ and $y$). We obtain the steady-state electromagnetic fields at the six boundaries of the simulation window, and compute the total emitted power $P$ by integrating the steady-state Poynting vector through them. At the $+z$ and $-z$ planes, we calculate overlap integrals of the radiated field with the field of the fundamental TE GaAs mode (Fig.~\ref{fig:Fig2}a left panel, at $\lambda=1100$~nm). This allows us to determine $\beta$, the fraction of the total emitted power that is carried through the $\pm z$ planes by the GaAs mode.

The mode transformer simulations are also performed with FDTD. We launch the fundamental TE GaAs mode of the waveguide structure in Fig.~\ref{fig:Fig1}d, shown in the left panel of Fig.~\ref{fig:Fig2}a, into the mode transformer, at the $z=0$ plane. We obtain the steady-state electromagnetic fields at the output ($z=20~\mum$) plane on the mode transformer, and calculate the overlap integral between this and the output $\SiN$ mode (right panel on Fig.~\ref{fig:Fig2}a). Dividing it by the launched input power we obtain the mode transformer coupling efficiency $\eta$.

We proceed similarly for the simulation of modal reflectivity and transmissivity for the photonic crystal reflector of Fig.~\ref{fig:Fig4}a. For reflectivity, we place a field monitor at the $z=0$ plane, and the source at $z=100$~nm.

To determine the mode volume $V_{eff}$ used in the Purcell factor estimate, we use $V_{eff}=\int_VdV\epsilon(\mathbf{r})\left|\mathbf{E}(\mathbf{r})\right|^2/\max\left\{\epsilon(\mathbf{r})\left|\mathbf{E}(\mathbf{r})\right|^2\right\}$,
where the volume integral is evaluated over the entire microring resonator. Because the ring radius is large ($R=10\mum$), we assume the whispering gallery mode fields across the microring cross-section have the same distribution as the fundametal TE GaAs mode of Fig.~\ref{fig:Fig2}a's left panel, and an azimuthal dependence $\exp(i\cdot m\phi)$. Then, $V_{eff}=2\pi\cdot R\cdot\int_AdA\epsilon(\mathbf{r})\left|\mathbf{E}(\mathbf{r})\right|^2/\max\left\{\epsilon(\mathbf{r})\left|\mathbf{E}(\mathbf{r})\right|^2\right\}$,
where $A$ is the cross-sectional waveguide area. The maximum Purcell factor (assuming spatial and polarization alignment of the dipole) is calculated with the expression $F_p=(3/4\pi^2)\cdot Q/V'_{eff}$, where $V'_{eff}$ is the mode volume in cubic wavelength in the GaAs.

\noindent \small{\textbf{Experimental determination of mode transformer coupling efficiency}}
Power transmission and reflection spectra $T_\text{ dev}$ and $R_\text{ dev}$ are determined experimentally using the setup in Fig.~S1a of the SI. Light from a fiber-coupled supercontinuum laser source is passed through a 3~dB fiber directional coupler and polarization controller, then launched into the input waveguide with a lensed fiber. Transmitted light is collected with another lensed fiber aligned to the output waveguide facet at the opposite edge of the chip, and sent to an optical spectrum analyzer (OSA). Reflected light is captured by the input fiber, and routed to the OSA via the 3~dB splitter.

To estimate a lower bound for $\eta$, we use a simple model to obtain an expression for the transmitted power at the output, $T_\text{ dev}$, as suggested in Fig.~\ref{fig:Fig4}a. Light launched at the input $\SiN$ waveguide is transferred with efficiency $\eta$ into the GaAs guide, whereas a residual ($1-\eta$) portion of the original power remains in the $\SiN$ guide.  Light transferred to the GaAs guide will be reflected with a reflectivity $R$ by the PhC, and transmitted through it with transmissivity $T$. The output mode transformer converts light transmitted through the PhC reflector back into the $\SiN$ guide, with efficiency $\eta$. We assume that the residual light that remains in the $\SiN$ after the input mode transformer is unaffected by the PhC, after which it is partially transferred with efficiency $\eta$ to the GaAs guide by the output mode converter, and is then lost as radiation at the terminated GaAs structure tip. Light collected by the output lensed fiber
thus has two components, one that remains in the $\SiN$ guide, and one that is transferred to and from the GaAs guide, and interacts with the PhC reflector. The maximum power collected by the output lensed fiber is $T_\text{ dev}$, with
\begin{equation}
T_\text{ dev}\leq\eta_\text{ ext}\left[\eta^2T+(1-\eta)^2+2\cdot\eta (1-\eta)\sqrt{T}\right].
\label{eq1}
\end{equation}
Inside the square brackets, the first and second terms correspond respectively to light transmitted through the PhC and residual light that remains in the $\SiN$ guide, and the third term comes from the interference between the two. The transmitted power for wavelengths in and out of the bandgap region are ${T_{dev,in}}$ and ${T_{dev,out}}$, respectively, and we define the extinction ratio $\alpha=\frac{T_{dev,in}}{T_{dev,out}}$. Because experimentally ${T_{dev,in}}$ is at least one order of magnitude lower than ${T_{dev,out}}$, we can assume that the PhC transmission at bandgap wavelengths is negligible, so that $T\approx0$ and
\begin{equation}
\alpha>\frac{\left(1-\eta\right)^2}{\eta^2T+(1-\eta)^2\pm2\cdot\eta (1-\eta)\sqrt{T}}>\frac{\left(1-\eta\right)^2}{\eta^2+(1-\eta)^2+2\cdot\eta (1-\eta)}
\end{equation}
Isolating $\eta$, we obtain the inequality $\eta^2+(2-\alpha)/(\alpha-1)+1<0$. The minimum root of the quadratic equation is our lower bound for $\eta$. For $\alpha=-20$~dB, as typically observed in our PhC spectra, $\eta>90$~\%, conservatively. For the peak extinction of $\approx25$~dB, $\eta>94$~\%.

\noindent \small{\textbf{Experimental determination of external coupling efficiency}}
The external coupling efficiency $\eta_\text{ ext}$ includes the chip-to-fiber coupling efficiency and propagation losses in the $\SiN$ waveguide leading to the device. We employ the setup of Fig.~S1a of the SI to obtain the transmitted power spectrum of a blank $\SiN$ waveguide (i.e., with no GaAs devices). Prior to this measurement, the polarization of the incident light is set to TE by probing a PhC reflector and minimizing the transmitted power over the photonic bandgap with the polarization controller. The lensed fibers are then aligned to the blank $\SiN$ waveguide, and the transmission spectrum is recorded. The spectrum is then normalized by the supercontinuum source power spectrum, obtained by bypassing the lensed fibers and the device. The resulting transfer function accounts for insertion losses through the two lensed fibers ($\approx31$~\%), and through the device, $IL_\text{ dev} = \eta_\text{ dev}^{-1} = \left(\eta_\text{ext,in}\cdot\eta_\text{ext,out}\right)^{-1}$.  Assuming that the waveguide facets are identical on both edges of the chip, $\eta_\text{ext,in}=\eta_\text{ext,out}=\eta_\text{ext}$, the external coupling efficiency is $\eta_\text{ext} = \sqrt{\eta_\text{dev}}$. Figure S1b in the SI shows the average measured $\eta_\text{ ext}$ for 3 different waveguides as a function of wavelength (the red curve and grey area correspond to the mean and standard deviation over the three measurements, respectively).
Averaging this curve across the 1100~nm to 1300~nm wavelength range produces $\eta_\text{ ext}=0.23 \pm 0.03$ (the uncertainty is obtained by propagating the standard deviations from the three devices). The theoretical mode-mismatch coupling efficiency is calculated with the overlap integral
\begin{equation}
\eta_\text{{facet}}=\frac{ \text{Re}\left\{
\iint_{S}(\mathbf{e}_f\times\mathbf{h}^*)\cdot\hat{z}\,dS \,
\iint_{S}(\mathbf{e}\times\mathbf{h}_f^*)\cdot\hat{z}\,dS\right\} }
{ \text{Re}\left\{
\iint_{S}(\mathbf{e}_f\times\mathbf{h}_f^*)\cdot\hat{z}\,dS \right\}
\, \text{Re}\left\{
\iint_{S}(\mathbf{e}\times\mathbf{h}^*)\cdot\hat{z}\,dS \right\} }
\label{eq:overlap}
\end{equation}
taken over the cross-sectional area $S$ of the input/output $\SiN$ waveguide. Here, $\mathbf{e}$ and $\mathbf{h}$ are
the electric and magnetic field components of the fundamental TE $\SiN$ input/output waveguide mode (right panel on Fig.~\ref{fig:Fig2}a), and $\mathbf{e}_f$ and $\mathbf{h}_f$ are the field components of a focused Gaussian beam with a spot size of $2.5~\mum$. The Gaussian beam spot size is consistent with specifications from the lensed fiber manufacturer. With eq.(\ref{eq:overlap}), we obtain $\eta_\text{facet}\approx26$~\% for a 580~nm thick $\times$ 800~nm wide $\SiN$ waveguide, at a wavelength of 1110~nm.

\noindent \small{\textbf{Second-order correlation measurements and fits}}
A Hanbury-Brown and Twiss (HBT) setup was used to obtain the second-order correlation function $g^{(2)}(\tau)$ of QD emission upon continuous-wave pumping. In our experiments, histograms of delays between detection events in the two single-photon detectors
were measured. We related these histograms to $g^{(2)}(\tau)$ as explained below. We first calculated delay probability distributions $C(\tau)$ by normalizing the delay histograms. Sufficiently far away from zero time delay, $C(\tau)\approx A\exp\left(-A\tau\right)$. We took the 1000 longest-delay bins of our histograms and perform a log-log linear fit to obtain $A$. The histograms were then normalized by $A$. For $\tau\approx0$, $g^{(2)}(\tau)\approx C(\tau)$ (see ref.~\onlinecite{verberk_photon_2003}).
The $g^{(2)}(\tau)$ data was modeled with the double-exponential function
\begin{equation}
g^{(2)}(\tau) = 1 + A_1\exp(\lambda_1\cdot \tau)+ A_2\exp(\lambda_2\cdot \tau),
\end{equation}
with $A_1+A_2 =-1$. This functional form is expected from a two-level system coupled to a single dark state~\cite{davanco_multiple_2014}, and describes both antibunching at $\tau=0$, bunching at some later time delay, and a return to the Poissonian level at $\tau\rightarrow\infty$ . To take into account the $\sigma\approx129$~ps time-response of detection system (see below for details), we convolved the $g^{(2)}(\tau)$ above with a Normal distribution function $N(\tau,\sigma)$:
\begin{equation}
g_C^{(2)}(\tau)=g^{(2)}(\tau)\ast N(\tau,\sigma)=1+A_1E_1(\tau)+A_2E_2(\tau),
\end{equation}
where
\begin{align}
E_n(\tau) =\frac{\lambda_n}{2}\exp\left(\frac{\lambda_n\sigma}{2}\right)
\left\{\erf\left(-\frac{\tau}{\sqrt{2}\sigma}+\frac{\lambda_n\sigma}{\sqrt{2}}\right)\text{e}^{-\lambda_n\tau}+ \right.\nonumber \\
\left. \erf\left(\frac{\tau}{\sqrt{2}\sigma}+\frac{\lambda_n\sigma}{\sqrt{2}}\right)\text{e}^{\lambda_n\tau}\right\}
\label{eq:g2conv}
\end{align}
and $n=1,2$. Finally, to account for a Poissonian background, we used~\cite{verberk_photon_2003}
\begin{equation}
g_{C,B}^{(2)}(\tau)=1+\frac{1}{(1+b)^2}\left[ g_C^{(2)}(\tau)\right].
\end{equation}
The fits shown in the main text were done using $g_{C,B}^{(2)}(\tau)$ above, through a nonlinear least-squares procedure. For the QD in a waveguide of Figs.~\ref{fig:Fig5}b and~\ref{fig:Fig5}c, the background $b$ was used as a fit parameter, while for the cavity-coupled QDs of Figs.~\ref{fig:Fig6}c and~\ref{fig:Fig6}e , $b$ was fixed at values estimated from fits to emission spectra (see below for spectrum fitting procedures). To plot $g^{(2)}(\tau)$ without the effect of the finite timing resolution, we used $\sigma=0$ in eq.(\ref{eq:g2conv}) and used the same fitting parameters. Uncertainties quoted for $g^{(2)}(0)$ are 95~\% fit confidence intervals, corresponding to 2 standard deviations.

\noindent \small{\textbf{Photoluminescence spectrum fits}}
The photoluminescence spectra in Figs.~\ref{fig:Fig6}b and ~\ref{fig:Fig6}f were fitted with a sum of three Lorentzians, representing
the cavity and two excitons, $X_1$ and $X_2$. A representative fitted spectrum is shown in Fig.~\ref{fig:Fig6}d, where the individual contributions are also displayed. To produce the left panel on Fig.~\ref{fig:Fig6}g, the different contributions were multiplied by a
spectrum representing the bandpass grating filter used experimentally, and the $X_1$ contribution was then normalized to the sum of the integrated intensities of all components before filtering. The wavelength detuning $\Delta$ between $X_1$ and the cavity was determined from these fits. All uncertainties quoted for $\Delta$ and the $X_1$, $X_2$ and cavity contributions correspond to 95~\% fit confidence intervals (two standard deviations).

\noindent \small{\textbf{Photoluminescence decay measurements}}
For excited state lifetime measurements, we employed a 10~GHz lithium niobate electro-optic modulator (EOM) to produce a 80 MHz,
$\approx200~$ps pulse train from the CW ECDL laser.
A fiber-based polarization controller (FPC) was used to control the polarization of the ECDL light going into
the EOM, and a DC bias was applied to the EOM to maximize signal extinction.
An electrical pulse source was used to produce an 80~MHz
train of $\approx200$~ps pulses of $< 1$~V peak amplitude, which was then amplified and used to drive the EOM via its radio frequency (RF) port.
A trigger signal from the pulse generator served as the reference channel in our TCSPC system.
Figure S3a in the SI shows a typical temporal profile for the pulses produced by the EOM, detected with an SNSPD. Pulse
FWHM of $\approx200$~ps and $> 20$~dB extinction are observed. The pulsed electrical signal produced small satellite peaks that were
imprinted in the optical signal, as indicated in Fig.~S3a. These satellite peaks typically appeared a few ns after each
proper pulse, and were $\approx20$~dB below the latter in intensity.
Impulse response functions (IRFs) such as the one in Fig.~S3a were used in decay lifetime fits as explained below, so
that the effect of satellite peaks, though minimal, was accounted for. Lastly, to determine the time resolution of our detection system, we launched attenuated few-picosecond pulses from a Ti:Sapphire
mode-locked laser at 975~nm into the SNSPDs, to obtain the temporal trace in Fig.~S3b. The peak can be well fitted
with a Gaussian with standard deviation $\sigma=129$~ps~$\pm~0.04$~ps (uncertainty is a 95~\% least-squares fit confidence interval,
corresponding to two standard deviations).

\noindent \small{\textbf{Photoluminescence decay fits}}
Quantum dot emission decay fits were performed using maximum likelihood estimation. We consider a lifetime trace $Y^k=\{Y_i\}_{i=1}^{k}$ where a known number of photon counts $N$ is distributed over $k$ time bins, such that the bin counts $y_i$ follow a multinomial distribution~\cite{kollner_how_1992}. The maximum likelihood estimator is
\begin{equation}
g_{MLE}(y^k) = \argmin_{\theta \in \Theta} \left\{- \sum_{i=1}^{k}y_i\ln p_i(\theta)\right\},
\label{eq:MLE}
\end{equation}
where $\theta$ is a vector in the multidimensional parameter space $\Theta$. Estimates for the various fit parameters are obtained by finding $\theta$ that minimizes the expression in the curly brackets, where $y_i$ is the $i$-th bin count, and $p_i(\theta)$ is a probability density function that models the decay, evaluated at the $i$-th bin. We define $p_i(\tau) = e^{-ir/k}\dfrac{e^{\frac{r}{k}} - 1}{1 - e^{-r}}$, with $r \triangleq \dfrac{i\cdot\Delta t}{\tau}$. For a monoexponential decay when a portion $b$ of the signal is due to background emission,
\begin{align}
p_i(\theta)=p_i(\tau,b) &= \dfrac{b}{k} + (1-b)p_i(\tau)
\end{align}
For biexponential decay with a background $b$, let $\tau \triangleq (\tau_1,\tau_2)^{T}$. Then $p_i(\tau,a,b)$ (where $a$ is the contribution of the first exponential decay) may be expressed as
\begin{align}
p_i(\theta)=p_i(\tau,b,a) &= \dfrac{b}{k} + (1-b)\left[ap_i(\tau_1) + (1 - a)p_i(\tau_2)\right]
\end{align}
Variances for the estimated parameters in
$\theta$ can be obtained from the diagonal elements of the inverse of the Fisher Information Matrix (see SI for further details). In the fitting procedure, the trial decay function $p_i(\theta)$ is numerically convolved with the experimentally measured, background-subtracted impulse response function (IRF) and used in eq.(\ref{eq:MLE}). Because the optical pulses used to obtain the IRF follow a considerably different path length towards the detector than the QD signal, the IRF and QD decay traces are delayed with respect to each other. We manually align the two traces to minimize fit residuals. Uncertainties given in the text correspond to standard deviations for the various parameters, obtained from the diagonal elements of the inverse of the Fisher information matrix computed with the expectation values from the fit (corresponding to the Cram{\' e}r-Rao lower bound).

\noindent \small{\textbf{Estimate of $\beta$}}
Below we estimate the coupling $\beta$ of the QD exciton at $\lambda\approx1330.18$~nm of Fig.~\ref{fig:Fig5}a into the guided TE mode of the GaAs waveguide where it was hosted. Ideally such a measurement would involve saturating the QD under pulsed excitation, where the maximum possible photon flux from the QD is given by the laser repetition rate.  Because a pulsed source with sufficient power to saturate the QD was unavailable, our estimate relied on the continuous-wave emission spectrum of Fig.~\ref{fig:Fig4}a. A three-level system model for the QD was then used to account for blinking. First, we measured the spectrum of a laser signal of known power at 1070~nm with our spectrometer, using the same fiber-coupled input as that for Fig.~\ref{fig:Fig5}a. The laser was attenuated with a calibrated variable optical attenuator (VOA), and launched into a fiber-based 10:90 power splitter (with a calibrated power-splitting ratio), the 90~\% port of which was sent to a photodiode for power monitoring. Integration of the background-subtracted laser spectrum counts divided by the laser power gave a factor of 0.0023 counts per photon at the spectrometer fiber-coupled input (this includes losses at the fiber connector, spectrometer slit, grating and output slit before the InGaAs detector array). This allowed us to obtain, from the fitted QD spectrum of Fig.~\ref{fig:Fig5}a, a photon flux $P=3.0\times10^6$~s$^{-1} \pm 0.5\times10^6$~s$^{-1}$ (errors come from the 95~\% fit confidence intervals) at this fiber input for the 1130.18~nm exciton line (accounting for the wavelength difference). We next expanded the photon flux as $P = X\beta\eta\eta_\text{ext.}\eta_\text{TF}$, where $X$ is the exciton population probability, $\eta$ the mode transformer efficiency, $\eta_\text{ext.}$ the lensed fiber-to-chip coupling efficiency,  and $\eta_\text{TF}=0.91 \pm 0.03$ is the lensed fiber transmission (uncertainty from measurement error, corresponding to one standard deviation). Solving the three-level system rate equations (with one bright and one dark transition) that fit the $g^{(2)}(\tau)$ data in Fig.~\ref{fig:Fig4}c - assuming the lifetime in Fig.~\ref{fig:Fig4}d for the bright transition - we obtain $X=0.15 \pm 0.04$, where the uncertainty is the 95~\% fit confidence interval. We note that connecting the dark state to either the ground or bright excited state in our model leads to $X\approx0.15$. Assuming $\eta=98~\%$ (the maximum from simulation) and $\eta_{ext.}=0.22$, a reasonable value from Fig.~S1b at 1130~nm, we obtain, propagating uncertainties, $\beta=0.20 \pm 0.07$.


\noindent \textbf{Acknowledgements}
We thank Daron Westly and Rob Ilic from the NIST CNST for invaluable aid with fabrication. J.L. acknowledges
support under the Cooperative Research Agreement between the
University of Maryland and NIST-CNST, Award 70NANB10H193.

\newpage
\onecolumngrid \bigskip

\begin{center} {{\bf \large SUPPLEMENTARY INFORMATION}}\end{center}

\setcounter{figure}{0}
\makeatletter
\renewcommand{\thefigure}{S\@arabic\c@figure}

\setcounter{equation}{0}
\makeatletter
\renewcommand{\theequation}{S\@arabic\c@equation}

\section{Extended discussion: quantum photonic integrated circuits with quantum dots}
The considerable potential of InAs/GaAs quantum dots (QDs), both for triggered single-photon generation and as quantum logic elements, has spurred the development of a number of platforms that seek to incorporate these QDs within photonic circuits.  A direct method is to develop both active and passive components within the same material system. Along these lines, monolithic GaAs-based quantum photonic circuits with on-chip quantum dot-based single-photon sources have been demonstrated by a number of research groups~\cite{dietrich_gaas_2016}. Two general approaches have been adopted. In the first, passive circuits are composed of low-index-contrast GaAs/AlGaAs ridge waveguides produced on top of a GaAs/AlGaAs substrate~\cite{jons_monolithic_2015,reithmaier_-chip_2015}. Such waveguide geometries can be produced with relatively straightforward fabrication processes, relying solely on epitaxial growth for vertical optical confinement and a single etching step for lateral confinement. Due to the small vertical refractive index contrast that is achievable through growth, waveguide cross-section dimensions are typically of the order of microns. While lower refractive indices are generally desirable for minimizing scattering losses in propagation, large mode field diameters (and large mode volumes in the case of cavities) translate into less compact devices, less effective light-matter interactions, and less effective geometrical control of waveguide dispersion (relevant for on-chip nonlinear optics). In particular, weak vertical confinement means more effective dipolar coupling to substrate radiative modes, translating into a reduced $\beta$-factor for radiating dipoles (see main text). Distributed-feedback reflector-based geometries such as in ref.~\onlinecite{jons_monolithic_2015} can ameliorate this, however the achievable mode-field diameters are also limited by the relatively small index contrast between GaAs and AlGaAs. All of these issues are to great extent circumvented in the second approach, in which circuits composed of suspended GaAs waveguides surrounded by air or vacuum, either of the channel~\cite{prtljaga_monolithic_2014} or photonic crystal~\cite{poor_efficient_2013} types (or both), are implemented. In this case, the strong index contrast allows strong transverse field confinement in waveguides of cross-sectional dimensions of the order of hundreds of nanometers. Small modal areas and cavity mode volumes can be achieved, meaning stronger light-matter interactions and higher $\beta$-factors for radiating dipoles~\cite{ref:Davanco2}, together with strong geometry-based control of waveguide dispersion. An important limitation of such an approach, however, is the losses due to scattering at the etched sidewalls, which can be considerably higher due to the strong index contrast between the semiconductor and the air. A second issue is that the fragility of suspended GaAs structures imposes limits on the dimensions of free-standing circuits, requiring support structures such as tethers or transitions to non-suspended waveguides, which may induce significant scattering losses~\cite{poor_efficient_2013}. Fabrication, device handling, and further integration with other types of on-chip elements are also more cumbersome in this case. An additional challenge common to both approaches is that passive circuits are produced in the same material layer that contains the QDs. Because there is no strict separation between active and passive portions of the photonic circuit, the population of QDs inside the passive section can contribute to excess optical absorption.

As discussed in the main text, our heterogeneous integration platform offers essentially all the advantages of the two approaches described above, while addressing many of the aforementioned challenges.  The large refractive index contrast between GaAs and Si$_3$N$_4$ allows for strong modal confinement within the GaAs layer so that large $\beta$ factors can be achieved in the active quantum dot region.  This large refractive index contrast is, in addition, achieved without requiring devices to be undercut, improving the mechanical and thermal stability of the system, particularly as the number of integrated elements increases.  Furthermore, complete removal of the GaAs material outside in the passive regions avoids excess optical absorption due to the background QD ensemble.  Within the passive sections, the large refractive index contrast between Si$_3$N$_4$ and SiO$_2$ enables the dispersion engineering and large effective nonlinearity needed for nonlinear optics applications, such as frequency downconversion of the QD emission to the 1550~nm telecom band~\cite{li_efficient_2016}.  Such nonlinear optics applications can in principle be implemented in suspended GaAs photonic circuits, though the much wider bandgap of Si$_3$N$_4$ and SiO$_2$ in comparison to GaAs-based materials ensures that two-photon absorption, an important factor in nonlinear nanophotonic devices, is negligible over a wide range of wavelengths.  Coupling off-chip to optical fibers can also be optimized, as our platform is compatible with end-fire approaches that utilize inverse tapers and symmetric low-index claddings~\cite{shoji_low_2002,tsuchizawa_microphotonics_2005}. In particular, devices can be designed to admit a full SiO$_2$ cladding rather than the current top air cladding, with additional processing likely consisting of a single additional PECVD deposition step.

While all of the fabrication steps that we have demonstrated in this work are scalable, the random nature of the in-plane spatial locations of self-assembled InAs/GaAs QDs is a limitation on the overall yield and ability to, for example, integrate multiple QD sites together within a passive Si$_3$N$_4$ circuit.  Going forward, we note that developments in site-controlled QD growth~\cite{rigal_site-controlled_2015,schneider_gaas/gaas_2012,helfrich_growth_2012} will help to address this issue, and in general will be compatible with our fabrication approach.  In the near-term, we can envision pre-characterization of QDs after creating the bonded wafer stack, such that GaAs devices are only created in regions for which desirable single QD behavior has been confirmed.  In particular, photoluminescence imaging~\cite{sapienza_nanoscale_2015} has been confirmed as a technique for locating the position of single InAs/GaAs QDs with respect to alignment features, and recent implementations~\cite{Jin_in_prep} have demonstrated the location of single QDs within $\approx$~5000~$\mu$m$^2$ spatial regions with sub-10~nm positioning uncertainty, with typical image acquisition times of 1~s.  We anticipate that such an approach can enable a higher throughput than pick-and-place techniques~\cite{zadeh_deterministic_2016,mouradian_scalable_2015,murray_quantum_2015,bermudez-urena_coupling_2015}, which also require pre-screening of the quantum emitters along with the additional assembly steps.

\newpage
\section{Fabrication Details}
To produce the starting wafer stack shown in Fig.~3a of the main text, we utilized the low-temperature plasma-activated direct wafer bonding procedure of ref.~\onlinecite{fang_hybrid_2007}. The layer stack for the two wafers that are bonded in this procedure, one silicon-based and one GaAs-based,
are given in tables~\ref{table:SiN} and~\ref{table:epi} respectively. The $\SiN$ layer of the silicon-based stack in table~\ref{table:SiN}
was grown with low-pressure chemical vapor deposition, and the epilayer stack of table~\ref{table:epi} was grown via molecular beam epitaxy.

\begin{table}[h]
\caption{$\SiN$ wafer stack}
\centering
\begin{tabular}{c c c}
\hline\hline
Layer & Material & Thickness (nm) \\ [0.5ex]
\hline
Waveguide & $\SiN$ & 550 \\
Bottom cladding & Thermal SiO$_2$ & 3000 \\
Substrate & Si & - \\ [1ex]
\hline
\end{tabular}
\label{table:SiN}
\end{table}

\begin{table}[h]
\caption{GaAs Epilayer Stack}
\centering
\begin{tabular}{c c c}
\hline\hline
Layer & Material & Thickness (nm) \\ [0.5ex]
\hline
Surface cap & GaAs & 10 \\
Waveguide top & Al$_{0.30}$Ga$_{0.70}$As & 40 \\
Waveguide top & GaAs & 74 \\
Quantum well & In$_{0.15}$Ga$_{0.85}$As & 6 \\
Quantum dot & InAs & 2.4 monolayer \\
Barrier & In$_{0.15}$Ga$_{0.85}$As & 1 \\
Waveguide bottom & GaAs & 74 \\
Sacrificial layer &  Al$_{0.30}$Ga$_{0.70}$As & 50 \\
Sacrificial layer &  Al$_{0.70}$Ga$_{0.30}$As & 1500 \\
Substrate & GaAs & - \\ [1ex]
\hline
\end{tabular}
\label{table:epi}
\end{table}

A $\approx30$~nm layer of SiN was deposited on top of cleaved ($\approx5$~mm$^2$) pieces of the III-V epiwafer with plasma-enhanced chemical vapor deposition (PECVD).
Contact lithography followed by reactive ion etching in a CHF$_3$/O$_2$/Ar plasma was used to produce $\approx10\mum$ wide, $\approx1$~cm long, $\approx30$~nm deep
channels on the $\SiN$ wafer surface, prior to bonding. This was done to prevent the formation of trapped H$_2$ bubbles at the
bonding interface during the annealing process~\cite{fang_hybrid_2007}. The $\SiN$ wafer was then cleaved into small ($\approx2$~cm$^2$) pieces.

For wafer bonding, the surfaces of the GaAs and $\SiN$ wafer pieces were cleaned in acetone,
then activated in an  O$_2$ plasma for 1 minute, at a pressure of 26.7 Pa (200~mTorr), flow of $1.5\times10^{-5}$ O$_2$ mol/s (20~sccm) and 200~W radio-frequency power. Pairs of wafers were then
placed in contact and pre-bonded under light manual contact. The pre-bonded samples were next annealed at 300~$^\circ$C for 1 hour in a nitrogen-purged
environment to produce a permanent bond. The warm-up and cool-down rates were set to 5~$^\circ$C/min. At this point, samples consisted of
small, rectangular-shaped GaAs wafer pieces permanently bonded onto the surface of larger $\SiN$ wafers.

We next carefully covered the exposed $\SiN$ areas on the wafers with Apiezon W wax~\cite{ref:NIST_disclaimer_note} that had been previously dissolved in trichloethylene (TCE). The dissolved wax wetted
the $\SiN$ surfaces and the sidewalls of the bonded GaAs pieces, however not the exposed back surfaces of the GaAs wafer. We placed the samples on a
hotplate at 80~$^\circ$C for 30 minutes to evaporate the TCE, solidifying the wax.
The samples were next immersed in a 3:7 H$_3$PO$_4$:H$_2$O$_2$
solution for approximately 5 hours, to remove most of the GaAs substrate. They were then transferred to a
4:1 citric acid (50~\% mass fraction):H$_2$O$_2$ solution, which etched GaAs with a very high selectivity with respect to the AlGaAs sacrificial layers.
The samples were left in for approximately 5 hours, until the exposed GaAs wafer surface looked uniform and unchanged.
At this point, the GaAs substrate had been completely removed.
Next, the samples were dipped in 49~\% HF for 30 seconds to remove the AlGaAs sacrifial layers. Finally, the wax was removed with TCE.

Following the wafer bonding step, fabrication proceeded as described in the main text. Further details are provided here. An array of Au alignment marks was first produced on top of the GaAs layer via electron-beam lithography followed by metal-lift-off.
A bilayer polymethylmethacrylate/copolymer resist process was used. An electron-beam evaporator was used to deposit a 10~nm Cr adhesion layer, and a 50~nm Au layer. Lift-off was carried out in an acetone bath.
Electron-beam lithography with ZEP 520A~\cite{ref:NIST_disclaimer_note} resist followed by inductively-coupled plasma etching using a Cl$_2$:Ar chemistry
were next used to define GaAs devices aligned to the Au mark array. Because ZEP520A is a positive-tone resist, devices were defined by
etching the GaAs only in micron-size areas that surrounded the devices. To remove the remaining GaAs from the rest of the wafer surface, we
used a wet-etch approach. First, e-beam lithography with ma-N 2045negative tone resist~\cite{ref:NIST_disclaimer_note} was performed to define protection patterns that covered
only the device areas and a selected number of Au alignment marks (a protection patch is highlighted in the optical micrograph Fig.~3b in the main text,
covering the GaAs microring resonator and bus waveguide). The samples were then immersed in TFA gold etchant for $\approx1$~min, then in 1020 Cr etch solution for $\approx1$~min. This removed exposed
Cr/Au alignment marks, as well as the exposed GaAs layer. The wet etch procedure could be repeated several times without affecting the
resist protection layer. Acetone was afterwards used to remove the ma-N resist.

After cleanup of the etched sample surface, a second electron-beam lithography exposure was performed, referenced to the original Au mark array,
to define $\SiN$ waveguide patterns aligned to the previously etched GaAs devices. We emphasize that the alignment marks used were from
the original mark array, and were protected during the GaAs wet etch step. Reactive ion etching (RIE) in a CHF$_3$/CF$_4$ plasma was
used to produce the $\SiN$ waveguides. The chip was finally cleaved perpendicular to the $\SiN$ waveguides $>1$~mm away from the GaAs devices, to allow access with optical fibers in the endfire configuration.

\newpage

\section{Room temperature photonic characterization setup}
\begin{figure}[!h]
\includegraphics[scale=2.0]{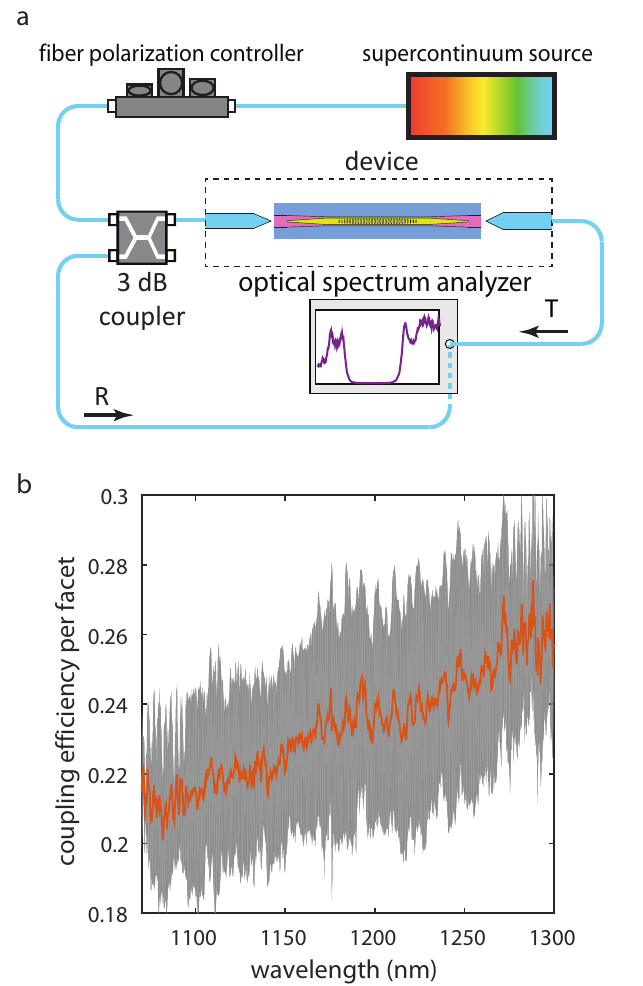}
\caption{\textbf{Photonic characterization setup}.\textbf{a}, Experimental setup for passive, endfire coupling measurement of transmitted (T) and reflected (R) power. \textbf{b}, Measured coupling efficiency per facet (including fiber-to-waveguide coupling and propagation losses in the waveguide between the facet and a GaAs device) as a function of wavelength. The red curve and grey area are the mean value and standard deviation over three measured waveguides.}
\label{SIfig:Fig1}
\end{figure}

\newpage

\section{Cryogenic measurement experimental setup }
\begin{figure}[!h]
\includegraphics[scale=1.25]{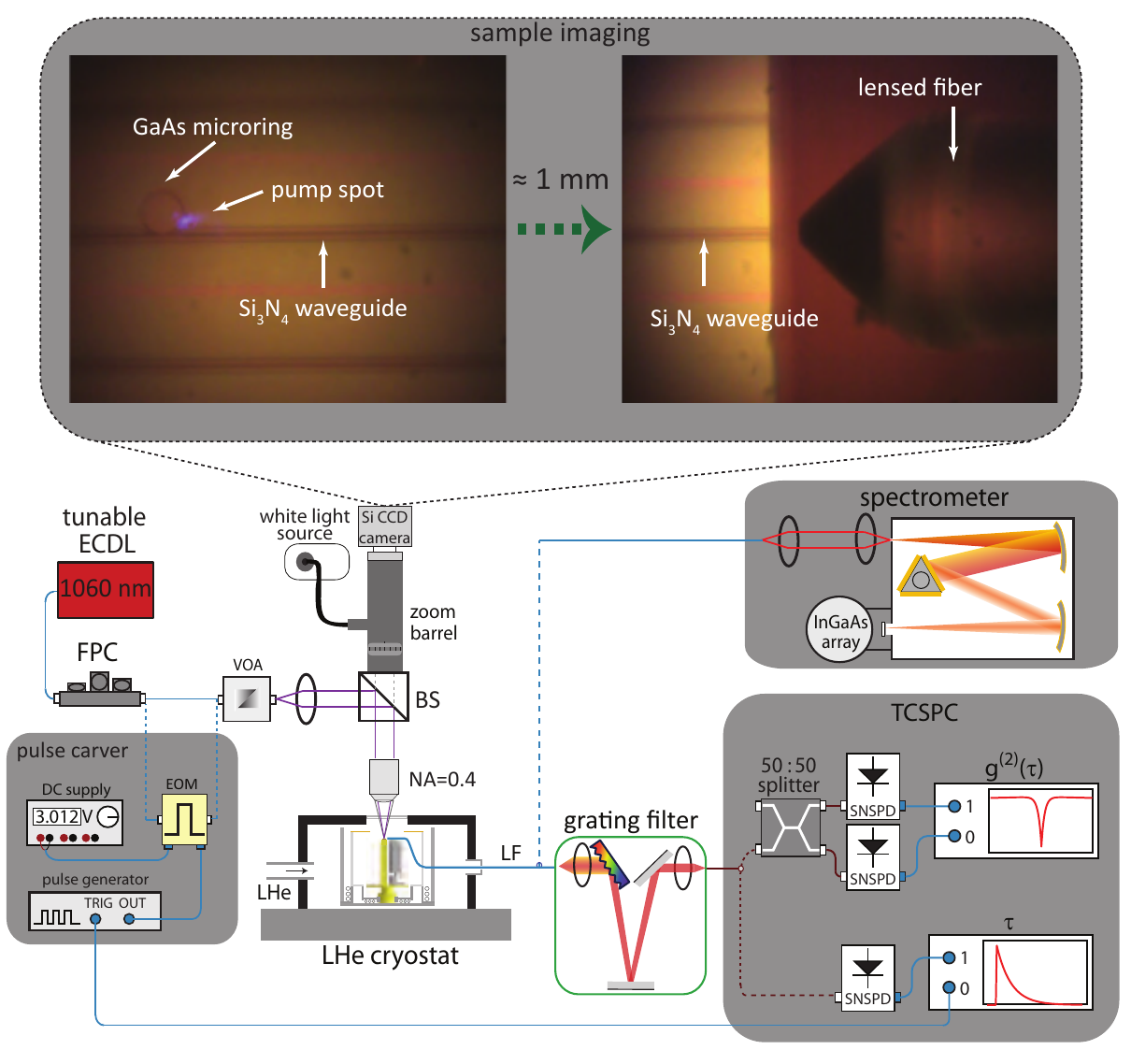}
\caption{\textbf{Cryogenic measurement setup}. Devices are tested inside a liquid Helium flow cryostat.
A window at the top allows imaging of the sample, as shown in the "sample imaging" box. Devices are also pumped from the top with laser light
(purple spot over portion of the GaAs microring resonator), and quantum dot photoluminescence is coupled into the $\SiN$ waveguide. The
emitted light travels $\approx1$~mm down the $\SiN$ waveguide until the cleaved edge of the chip, where it is collected with a lensed fiber
mounted to a nanopositioning stage inside the cryostat. The collected light can then be routed to a grating spectrometer, or filtered in a
grating filter and then routed towards time-correlated single-photon counting measurements. Pulsed pumping can be achieved by modulating
continuous-wave laser light with an electro-optic modulator.  ECDL: external-cavity diode laser; FPC: fiber polarization controller;
EOM: electro-optic modulator; BS: beamsplitter;
VOA: variable optical attenuator. LF: lensed fiber. SNSPD: superconducting nanowire single-photon detector;
TCSPC: time-correlated single-photon counting; CCD: charge-coupled device; LHe: liquid Helium}
\label{SIfig:Fig2}
\end{figure}

The setup used for low temperature cryogenic measurements is shown in Fig.~\ref{SIfig:Fig2}. Samples were placed on a fixed mount inside
a liquid Helium flow cryostat and cooled down $\approx7$~K. A microscope consisting of a long-working distance
objective (20$\times$, NA=0.4), beamsplitter (BS) and combination zoom lens / illuminator system mounted at the top cryostat
window allowed devices on the sample surface to be imaged. An example image of a GaAs microring resonator device can be seen
in the "sample imaging" box in the schematic of Fig.~\ref{SIfig:Fig2}. Light from an external cavity tunable laser (ECDL) with center
wavelength around 1060~nm was introduced into the objective via the beamsplitter, to produce a small ($\approx5~\mum$) spot that
pumped QDs at select locations on the device under test. As discussed in the main text, photoluminescence from the QDs was coupled
to $\SiN$ waveguides, and collected at the cleaved edge of the chip with a lensed fiber. A nanopositiong stage stack placed inside the cryostat allowed the lensed fibers to be aligned to waveguide facets, as illustrated in Fig.~\ref{SIfig:Fig2}.

\newpage

The collected photoluminescence was either be routed to a grating spectrometer equipped with a liquid-nitrogen cooled~InGaAs photodiode array,
or filtered through a $\approx700$~pm bandpass tunable grating filter and then routed towards our single-photon detection system for
time-correlated single-photon counting (TCSPC) measurements of excited state lifetime and second-order correlations. For correlation
measurements, we used a Hanbury-Brown and Twiss setup consisting of fiber-based 50:50 beamsplitter connected to two
amorphous WSi superconducting nanowire single-photon detectors (SNSPDs)~\cite{marsili_detecting_2013}. Detector counts were correlated in
a TCSPC unit.

\begin{figure}[!h]
\includegraphics[width=\linewidth]{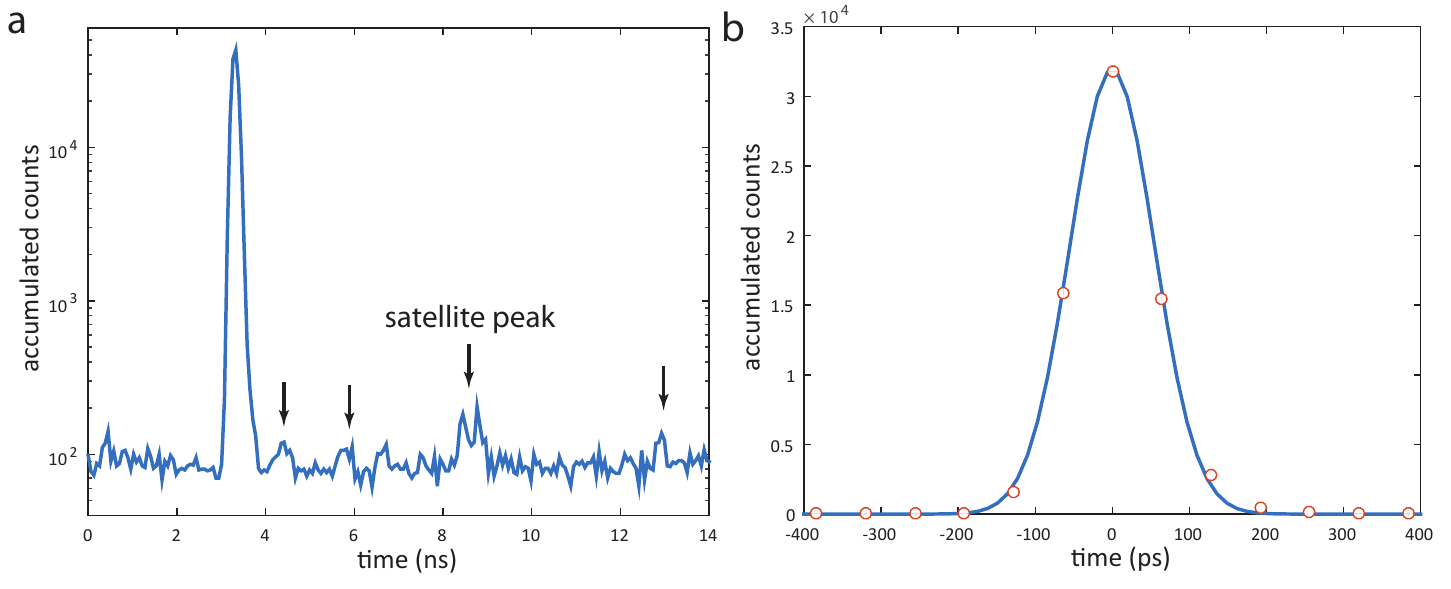}
\caption{\textbf{Impulse response functions}. \textbf{a}, Temporal profile of pulsed pump achieved with an electro-optic modulator
(EOM). Satellite peaks due to an imperfect electrical driving signal are pinpointed with arrows. \textbf{b}, Impulse response of the
SNSPD-based detection system, obtained with a $\approx$~ps pulsed Ti:Sapphire laser source.}
\label{SIfig:Fig3}
\end{figure}

For lifetime measurements, a 10~GHz lithium niobate electro-optic modulator (EOM) was used to produce a 80~MHz,
$\approx200~$ps pulse train from the CW ECDL laser~\cite{dada_indistinguishable_2016}.
A fiber-based polarization controller (FPC) was used to control the polarization of the ECDL light going into
the EOM, and a DC bias was applied to the EOM to maximize signal extinction.
An electrical pulsed source was used to produce an 80~MHz train of $\approx200$~ps pulses of $< 1$~V peak amplitude, which was then amplified and used to drive the EOM modular via its radio frequency (RF) port.
A trigger signal from the pulse generator served as the reference channel in our TCSPC system.
Figure~\ref{SIfig:Fig3}a shows a typical temporal profile for the pulses produced by the EOM, detected with an SNSPD. Pulse
FWHM of $\approx200$~ps and $> 20$~dB extinction are observed. The pulsed electrical signal produced small satellite peaks that were
imprinted in the optical signal, as indicated in Fig.~\ref{SIfig:Fig3}a.
Impulse response functions (IRFs) such as shown in Fig.~\ref{SIfig:Fig3}a were used in decay lifetime fits as explained below, so
that the effect of satellite peaks, though minimal, was accounted for.

To determine the time resolution of our detection system, we launched attenuated, few-picosecond pulses from a Ti:Sapphire
mode-locked laser at 975~nm into the SNSPDs, to obtain the temporal trace in Fig.~\ref{SIfig:Fig3}. The peak can be well fitted
with a Gaussian with standard deviation $\sigma=129$~ps~$\pm~0.04$~ps (uncertainty is a 95~\% least-squares fit confidence interval,
corresponding to two standard deviations).
\newpage
\section{Purcell enhancement in the heterogeneous platform}
\begin{figure}[!h]
\begin{center} \includegraphics[scale = 1.65]{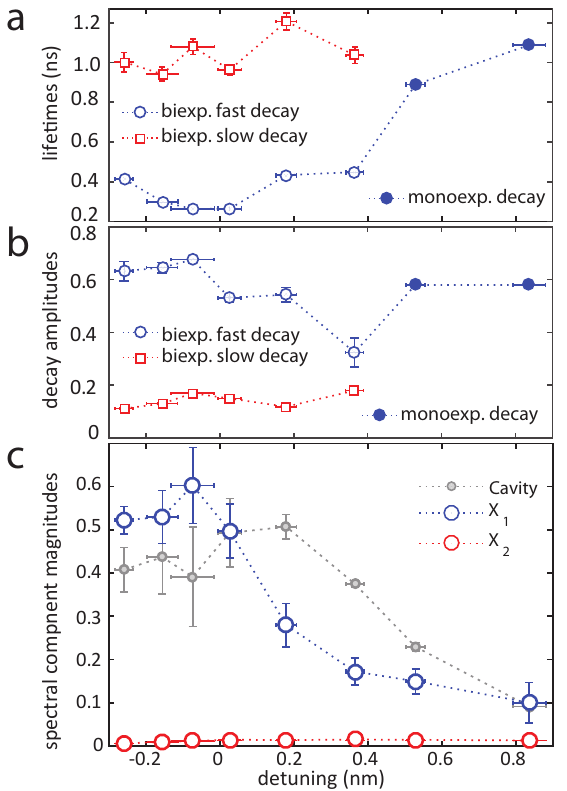} \caption{\textbf{Lifetime and PL intensity fit
results}. \textbf{a}, Decay lifetimes for the fits in Fig.~6f, as a function of the detuning $\Delta$. Open blue circles and
red squares are the fast and slow biexponential decay lifetimes, closed blue circles are the monoexponential decay lifetimes. \textbf{b},
Corresponding amplitudes for the different decay components. \textbf{c}, magnitudes of the X$_1$, X$_2$
and cavity Lorentzians used to fit the PL spectra in \textbf{f}, normalized to the sum of the integrated intensities of the three components before filtering.
Uncertainties for $\Delta$ and the $X_1$, $X_2$ and cavity PL contributions are 95~\% fit
confidence intervals (two standard deviations). Uncertainties for $\tau$ are single standard deviations from the exponential decay fit
procedure.  } \label{SIfig:Fig5}
\end{center} \end{figure}

Here we provide further information about the Purcell enhancement data and fits for the quantum dot in a microring resonator presented in
Figs.~6d-6g of the main text. Figures~\ref{SIfig:Fig5}a and ~\ref{SIfig:Fig5}b respectively show lifetimes and corresponding decay
component amplitudes for the fits, as a function of the detuning $\Delta$. It is apparent that the fast lifetimes  vary considerably,
from $\approx414$~ps at $\Delta\approx-0.26$~nm to $\approx263$~ps at $\Delta\approx-0.07$~nm, then to $\approx~1.1$~ns at
$\Delta=0.84$~nm. Slow lifetimes remain consistently above 1~ns. The fast decay contribution remains above 50~\% for all detunings
except $\Delta\approx0.36$~nm, where a second exciton ($X_2$) is seen to couple to the same whispering-gallery mode in Fig.~6f.
The contributions of the two excitons and the cavity to the detected signal in the lifetime measurements is estimated through fits to the
emission spectra at each detuning, shown in Fig.~\ref{SIfig:Fig5}c. The $X_1$ contribution is seen to be dominant everywhere
(except $\Delta\approx0.36$~nm). The $X_2$ contribution is maximized at $\Delta\approx0.36$~nm, but remains below 0.015 everywhere.
These results indicate that the fast lifetimes can be assigned to the $X_1$ exciton. Further supporting this assignment is the fact
that the good quality of the $g^{(2)}(\tau)$ fit in Fig.~6e was achieved by including a Poissonian background equal to
the cavity contribution in the PL spectrum fit of Fig.~6d.

Uncertainties for $g^{(2)}(0)$, $\Delta$
and the $X_1$, $X_2$ and cavity PL contributions are 95~\% fit confidence intervals (two standard deviations). Uncertainties for
$\tau$ are single standard deviations from the exponential decay fit procedure.

\newpage
\section{Fisher information matrix}
Variances for the estimated decay lifetime parameters in $\theta$ can be obtained from the diagonal elements of the inverse of the Fisher Information Matrix $M(\theta)$ with
\begin{align}
\left[M(\theta)\right]_{mn} = \mathbb{E}_\theta\left[\dfrac{\partial \ln f^{(k)}_\theta(Y^k)}{\partial\theta_m}\dfrac{\partial \ln f^{(k)}_\theta(Y^k)}{\partial\theta_n} \right].
\end{align}
Here, $\mathbb{E}_\theta$ stands for expectation value, and $f^{(k)}_\theta(y^k)$ is the probability mass function for the counts in each bin of
the liftetime trace, which constitute a sequence of random variables with multinomial distibution~\cite{kollner_how_1992}:
\begin{equation}
f^{(k)}_\theta(y^k) = \dfrac{N!}{\prod_{i=1}^{k}y_i!}\prod_{i=1}^{k}p_i^{y_i}(\theta),~ \sum_{i=1}^{k}y_i = N, ~\sum_{i=1}^{k}p_i(\theta) = 1
\end{equation}
Noting that
\begin{equation}
\dfrac{\partial \ln f^{(k)}_\theta(y^k)}{\partial\theta_m} = \sum_{i=1}^{k}\dfrac{y_i}{p_i(\theta)}\dfrac{\partial p_i(\theta)}{\partial\theta_m},
\end{equation}
we can write
\begin{align}
\sum_{i=1}^{k}\sum_{j=1}^{k}\dfrac{\mathbb{E}_\theta\left[Y_iY_j\right]}{p_i(\theta)p_j(\theta)}\dfrac{\partial p_i(\theta)}{\partial \theta_m}\dfrac{\partial p_j(\theta)}{\partial \theta_n}
\label{eq:five}
\end{align}
Now
\begin{equation}
\mathbb{E}_\theta\left[Y_iY_j\right] =
\left\{
	\begin{array}{ll}
		\mbox{var}_\theta\left[Y_i\right] + \left(\mathbb{E}_\theta\left[Y_i\right]\right)^2 = (N^2 - N)p^2_i(\theta) + Np_i(\theta)   & \mbox{if } i = j, \\
		\mbox{COV}_\theta\left[Y_i, Y_j\right] + \mathbb{E}_\theta\left[Y_i\right]\mathbb{E}_\theta\left[Y_j\right] = (N^2 - N)p_i(\theta)p_j(\theta) & \mbox{if } i \neq j,
	\end{array}
\right.
\label{eq:corr}
\end{equation}
where $\mbox{var}_\theta$ and $\mbox{COV}_\theta$ are the variance and the covariance operators with respect to $\theta$, respectively. Substituting~(\ref{eq:corr}) into~(\ref{eq:five}), it follows that
\begin{equation}
\begin{aligned}
\mathbb{E}_\theta\left[\dfrac{\partial \ln f^{(k)}_\theta(Y^k)}{\partial\theta_m}\dfrac{\partial \ln f^{(k)}_\theta(Y^k)}{\partial\theta_n} \right] =& \sum_{i=1}^{k}\dfrac{(N^2 - N)p^2_i(\theta) + Np_i(\theta)}{p^2_i(\theta)}\dfrac{\partial p_i(\theta)}{\partial \theta_m}\dfrac{\partial p_i(\theta)}{\partial \theta_n} \\
& + \underbrace{\sum_{i=1}^{k}\sum_{j=1}^{k}}_{i \neq j}\dfrac{(N^2 - N)p_i(\theta)p_j(\theta)}{p_i(\theta)p_j(\theta)}\dfrac{\partial p_i(\theta)}{\partial \theta_m}\dfrac{\partial p_j(\theta)}{\partial \theta_n}
\end{aligned}
\label{eq:seven}
\end{equation}
The first summation can be written as
\begin{align}
\sum_{i=1}^{k}\dfrac{(N^2 - N)p^2_i(\theta) + Np_i(\theta)}{p^2_i(\theta)}\dfrac{\partial p_i(\theta)}{\partial \theta_m}\dfrac{\partial p_i(\theta)}{\partial \theta_n} = (N^2 - N)\sum_{i=1}^{k}\dfrac{\partial p_i(\theta)}{\partial \theta_m}\dfrac{\partial p_i(\theta)}{\partial \theta_n} + N\sum_{i=1}^{k}\dfrac{1}{p_i(\theta)}\dfrac{\partial p_i(\theta)}{\partial \theta_m}\dfrac{\partial p_i(\theta)}{\partial \theta_n}.
\label{eq:first_sum}
\end{align}
The second summation can be written as
\begin{align}
\underbrace{\sum_{i=1}^{k}\sum_{j=1}^{k}}_{i \neq j}\dfrac{(N^2 - N)p_i(\theta)p_j(\theta)}{p_i(\theta)p_j(\theta)}\dfrac{\partial p_i(\theta)}{\partial \theta_m}\dfrac{\partial p_j(\theta)}{\partial \theta_n} = (N^2 - N)\underbrace{\sum_{i=1}^{k}\sum_{j=1}^{k}}_{i \neq j}\dfrac{\partial p_i(\theta)}{\partial \theta_m}\dfrac{\partial p_j(\theta)}{\partial \theta_n}
\label{eq:second_sum}
\end{align}
Substituting~(\ref{eq:first_sum}) and~(\ref{eq:second_sum}) into~(\ref{eq:seven}), it follows that
\begin{align}
\mathbb{E}_\theta\left[\dfrac{\partial \ln f^{(k)}_\theta(Y^k)}{\partial\theta_m}\dfrac{\partial \ln f^{(k)}_\theta(Y^k)}{\partial\theta_n} \right] =
(N^2 - N)\sum_{i=1}^{k}\sum_{j=1}^{k}\dfrac{\partial p_i(\theta)}{\partial \theta_m}\dfrac{\partial p_j(\theta)}{\partial \theta_n} + N\sum_{i=1}^{k}\dfrac{1}{p_i(\theta)}\dfrac{\partial p_i(\theta)}{\partial \theta_m}\dfrac{\partial p_i(\theta)}{\partial \theta_n}.
\end{align}
Noting that
\begin{align}
\sum_{i=1}^{k}\sum_{j=1}^{k}\dfrac{\partial p_i(\theta)}{\partial \theta_m}\dfrac{\partial p_j(\theta)}{\partial \theta_n} = \sum_{i=1}^{k}\dfrac{\partial p_i(\theta)}{\partial \theta_m}\sum_{j=1}^{k}\dfrac{\partial p_j(\theta)}{\partial \theta_n} = \dfrac{\partial \left(\sum_{i=1}^{k} p_i(\theta)\right)}{\partial \theta_m}\dfrac{\partial\left(\sum_{j=1}^{k} p_j(\theta)\right)}{\partial \theta_n} = \dfrac{\partial (1)}{\partial \theta_m}\dfrac{\partial (1)}{\partial \theta_n} = 0,
\end{align}
it follows that
\begin{equation}
\mathbb{E}_\theta\left[\dfrac{\partial \ln f^{(k)}_\theta(Y^k)}{\partial\theta_m}\dfrac{\partial \ln f^{(k)}_\theta(Y^k)}{\partial\theta_n} \right] = N\sum_{i=1}^{k}\dfrac{1}{p_i(\theta)}\dfrac{\partial p_i(\theta)}{\partial \theta_m}\dfrac{\partial p_i(\theta)}{\partial \theta_n}.
\label{eq:Fisher_calc}
\end{equation}
We next define
\begin{align}
	p_i(\tau) & = e^{-ir/k}\dfrac{e^{\frac{r}{k}} - 1}{1 - e^{-r}}, ~r \triangleq \dfrac{i\cdot\Delta t}{\tau}\\
\dfrac{\partial p_i(\tau)}{\partial \tau} & = \dfrac{-r}{\tau} p_i(\tau)\left(-\dfrac{i}{k} + \dfrac{e^{r/k}}{(e^{r/k} - 1)k} - \dfrac{e^{-r}}{1 - e^{-r}}\right).
\end{align}
For a monoexponential decay when a portion $b$ of the signal is due to background emission,
\begin{align}
p_i(\tau,b) &= \dfrac{b}{k} + (1-b)p_i(\tau)
\end{align}
The Fisher matrix in this case can be computed with eq.(\ref{eq:Fisher_calc}) and
\begin{align}
\dfrac{\partial p_{i}(\tau,b)}{\partial b} &= \dfrac{1}{k} - p_i(\tau)\\
\dfrac{\partial p_i(\tau,b)}{\partial \tau} & = (1-b)\dfrac{-r}{\tau} p_i(\tau)\left(-\dfrac{i}{k} + \dfrac{e^{r/k}}{(e^{r/k} - 1)k} - \dfrac{e^{-r}}{1 - e^{-r}}\right)
\end{align}
For biexponential decay with a background, let $\tau \triangleq (\tau_1,\tau_2)^{T}$. Then $p_i(\tau,a,b)$ (where $a$ is the contribution of the first exponential decay) may be expressed as
\begin{align}
p_i(\tau,b,a) &= \dfrac{b}{k} + (1-b)\left[ap_i(\tau_1) + (1 - a)p_i(\tau_2)\right]
\end{align}
The Fisher matrix in this case can be computed with eq.(\ref{eq:Fisher_calc}) and
\begin{align}
\dfrac{\partial p_i(\tau,b,a)}{\partial b} &= \dfrac{1}{k} - \left[ap_i(\tau_1) + (1-a)p_i(\tau_2)\right]\\
\dfrac{\partial p_i(\tau,b,a)}{\partial \tau_1} & = a(1-b)\dfrac{\partial p_i(\tau_1)}{\partial \tau_1} \\
\dfrac{\partial p_i(\tau,b,a)}{\partial \tau_2} & = (1-a)(1-b)\dfrac{\partial p_i(\tau_2)}{\partial \tau_2} \\
\dfrac{\partial p_i(\tau,b,a)}{\partial a} & = (1-b)\left(p_i(\tau_1) - p_i(\tau_2)\right).
\end{align}
\newpage
\section{Optimized dipole coupling into the hybrid waveguide}
\begin{figure}[!h]
\includegraphics[width=\linewidth]{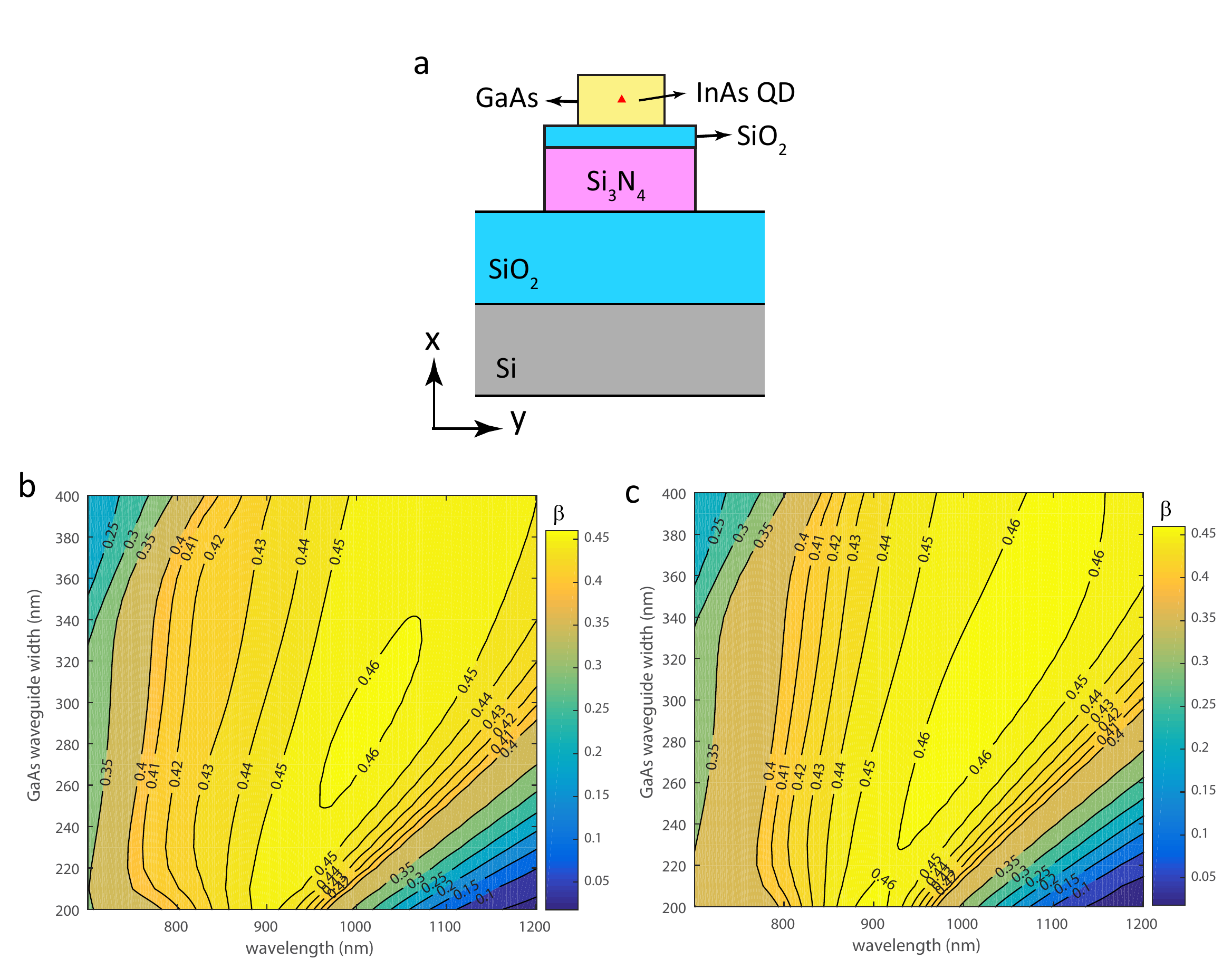}
\caption{\textbf{$\beta$-factors in optimized structures}. \textbf{a} Emission capture structure cross-section. \textbf{b} $\beta$ factors as a function of wavelength and GaAs waveguide width for the fundamental TE GaAs mode, for a GaAs waveguide with thickness of 190~nm, $\SiN$ waveguide has a width of 600~nm and thickness of 550~nm. A 100~nm thick layer of SiO$_2$ separates the $\SiN$ and GaAs waveguides. The $\SiN$ dimensions here are relevant for optical nonlinearities, as discussed in the main text. \textbf{c}, same as \textbf{a} for a 250~nm $\SiN$ thickness, relevant for linear optics applications.}
\label{SIfig:Fig4}
\end{figure}

Here we present simulation results for the fundamental TE GaAs mode $\beta$-factor of two optimized emission capture structures.
In both cases, the active waveguide section is a as shown in Fig.~\ref{SIfig:Fig4}a. The GaAs waveguide has a thickness of 190~nm, and the $\SiN$ waveguide width is 600~nm. A 100~nm layer of SiO$_2$ separates the two waveguides. Such a layer
can be produced with PECVD, same as the nitride layer grown on our GaAs wafer prior to bonding, without adversely affecting the bond quality.
For the first optimized geometry, Fig.~\ref{SIfig:Fig4}b, the $\SiN$ thickness is 550~nm, similar to the $\SiN$ waveguides in our sample. In Fig.~\ref{SIfig:Fig4}b,
the $\SiN$ thickness is 250~nm. In both cases, $\beta>0.46$ for modes propagating in either $+z$ or $-z$ directions ($2\beta>0.92$ altogether),
for a wavelength range of tens of nanometers around 1100~nm, for GaAs waveguide widths close to 300~nm.
\newpage

\begin{thebibliography}{10}
\bibitem{ref:NIST_disclaimer_note}
The identification of any commercial product or trade name is used to foster
  understanding. Such identification does not imply recommendation or
  endorsement or by the National Institute of Standards and Technology, nor
  does it imply that the materials or equipment identified are necessarily the
  best available for the purpose.
\end{thebibliography}

\end{document}